\documentclass[amsmath,amssymb,reprint,aps,prd,eqsecnum,unsortedaddress]{revtex4-2}
 
\usepackage{graphicx}
\usepackage{dcolumn}
\usepackage{bm}
\usepackage{mathrsfs}
\usepackage{todonotes}
\usepackage{stmaryrd}

\newcommand{\subAH}{\textrm{\textit{\tiny{A,H}}}}
\newcommand{\subAB}{\textrm{\textit{\tiny{A,B}}}}
\allowdisplaybreaks

\begin{document}

\title{Quantum effects of charged massive scalar fields on charged black hole space-times}

\author{Cormac Breen}

\email{cormac.breen@tudublin.ie}

\affiliation{School of Mathematics and Statistics,
Technological University Dublin,
Grangegorman,
Dublin 7, Ireland}

\author{George Montagnon}

\email{GJMontagnon@googlemail.com}

\affiliation{School of Mathematical and Physical Sciences,
University of Sheffield,
Hicks Building,
Hounsfield Road,
Sheffield. S3 7RH United Kingdom}

\author{Peter Taylor}

\email{peter.taylor@dcu.ie}

\affiliation{Centre for Astrophysics and Relativity,
School of Mathematical Sciences,
Dublin City University,
Glasnevin,
Dublin 9, Ireland.}

\author{Elizabeth Winstanley}

\email{E.Winstanley@sheffield.ac.uk}

\affiliation{School of Mathematical and Physical Sciences,
University of Sheffield,
Hicks Building,
Hounsfield Road,
Sheffield. S3 7RH United Kingdom}

\date{\today}

\begin{abstract}
We compute the renormalized expectation values of the scalar condensate, charge current and stress-energy tensor for a massive charged quantum scalar field on a Reissner-Nordstr\"om black hole. The mass of the scalar field is sufficiently large that classical superradiance is absent. We present results for the Boulware, Unruh and Hartle-Hawking states.
Renormalized expectation values in the Hartle-Hawking state are computed using an efficient and accurate mode sum prescription, and for the other two states we employ state subtraction.
The efficiency of our numerical method enables us to explore a large part of the parameter space, including near-extremal black hole charges and scalar field masses close to the superradiant limit.   
\end{abstract}

\maketitle

\section{Introduction}
\label{sec:intro}

While there are several proposals for how to quantize space-time in different contexts, there is not yet a consensus on a general theory of quantum gravity. 
Nevertheless, there are scenarios of interest where one can employ uncontroversial and robust approximations to quantum gravity. For example, in the context of black holes, at least away from curvature singularities, the semiclassical approximation has proven to be a very useful tool for studying quantum effects in these space-times. 
One of the many profound results arising from this program is the discovery that black holes formed by gravitational collapse emit quantum thermal radiation \cite{Hawking:1975vcx}. 

In semiclassical gravity, the field equations for the metric of space-time take the following form, which we will call the semiclassical Einstein equations,
 \begin{equation}
 \label{eq:semieqs}
		G_{\alpha \beta }+\Lambda g_{\alpha \beta }+{\mathcal {C}}_{1} H^{(1)}_{\alpha \beta  }+{\mathcal {C}}_{2} H^{(2)}_{\alpha \beta }= 8\pi T_{\alpha \beta }^{{\mathrm {(cl)}}} +8\pi \langle\hat{T}_{\alpha \beta }\rangle ,
	\end{equation}
where, on the left-hand-side, $g_{\alpha \beta }$ is the metric tensor, $G_{\alpha \beta }$ is the Einstein tensor and $\Lambda $ the cosmological constant (here, and throughout this paper, we use units in which $G=c=\hbar =k_{\mathrm {B}}=1$). 
On the right-hand side, we separate the contributions to the total stress-energy into two parts, a contribution $\langle\hat{T}_{\alpha \beta }\rangle$ which is the renormalized expectation value of the stress-energy tensor (RSET) for the quantized fields in some quantum state, and a contribution $T_{\alpha \beta }^{{\mathrm {(cl)}}}$  which is the (classical) stress-energy tensor for any additional background matter sources which are not quantized. 
The tensors $H^{(1)}_{\alpha \beta }$, $H^{(2)}_{\alpha \beta }$ are conserved geometrical tensors that are quadratic in the curvature and arise through the point-splitting regularization process that yields $\langle\hat{T}_{\alpha \beta }\rangle_{}$ (see, for example, \cite{Decanini:2005eg}). 
This regularization process corresponds to an infinite renormalization of the constants $\Lambda$, ${\mathcal {C}}_{1}$ and ${\mathcal {C}}_{2}$ (as well as Newton's constant $G$ but we are working in units where this is unity).  
Non-perturbative solutions to Eq.~(\ref{eq:semieqs}) are very difficult to obtain. In fact, it is practically impossible to write down an explicit expression for the RSET without some \textit{a priori}, highly symmetrized, ansatz for the metric. Instead one usually solves the semiclassical equations by perturbing around some background classical solution. At linear order in this prescription, the RSET $\langle {\hat {T}}_{\alpha \beta } \rangle$ evaluated on a known fixed background space-time governs the backreaction of the quantum field on the background geometry. Thus being able to efficiently and accurately compute the RSET on a fixed black hole space-time is central to understanding their semiclassical evolution. 

In this paper, we are motivated by solving the semiclassical Einstein-Maxwell equations which couple the semi-classical gravitational equations (\ref{eq:semieqs}) with the semiclassical Maxwell equations
\begin{equation}
\nabla_{\beta }F^{\beta \alpha } = 4\pi \langle {\hat {J}}^{\alpha } \rangle ,
\label{eq:SCMaxwell}
\end{equation}
where $F^{\alpha \beta }=\nabla ^{\alpha }A^{\beta }-\nabla ^{\beta }A^{\alpha }$ is the Faraday tensor, $A^{\alpha }$ is the electromagnetic potential and here $\langle {\hat {J}}^{\alpha }\rangle $ is the renormalized expectation value of the current operator for the quantized fields (and we are using Gaussian units). As before, we can only hope to solve these equations perturbatively around an electrovacuum solution to the classical Einstein-Maxwell equations. Therefore, to obtain the solutions at linear order requires a method for computing the expectation value of both the current and the RSET on the fixed background.

In particular, we will consider a charged quantum scalar field propagating on the Reissner-Nordstr{\"o}m black hole space-time, which is the unique spherically-symmetric electrovac solution of the Einstein-Maxwell equations.    
The charge of the quantum field introduces an effective chemical potential into the Hawking radiation \cite{Gibbons:1975kk,Hawking:1975vcx}, thereby modifying the emission and extracting charge as well as mass from the black hole.
A charged quantum scalar field will therefore back-react not only on the background space-time, but also on the background electromagnetic field, since it provides
the source term in Eq.~(\ref{eq:SCMaxwell}).

While the formal renormalisation process is well understood (see for example \cite{Christensen:1976vb,Christensen:1978yd, Fulling:1978ht,Decanini:2005eg,Balakumar:2019djw}), the practical calculation of renormalized quantities remains highly nontrivial. 
The Green function for the quantum field is typically expressed as a mode sum, with individual modes obtained numerically, from which, in order to renormalise, we must subtract geometric quantities defined through local expansions. 
Achieving an efficient and accurate implementation of this procedure therefore represents a significant technical challenge (see \cite{Herman:1995hm,Herman:1998dz} for early work in this direction for a charged scalar field).

For this reason, the majority of work in the literature concerning expectation values of either the current or stress-energy tensor for a quantum charged scalar field has focused on quantities which have either vanishing or finite renormalization subtraction terms \cite{Balakumar:2020gli,Klein:2021les,Klein:2021ctt,Alberti:2025mpg} or the differences in expectation values between two quantum states \cite{Balakumar:2022yvx}.

In contrast, for a neutral scalar field, over the past decade methodology has been developed to enable the practical computation of {\em {all}} components of the RSET. 
The current state-of-the-art employs one of two implementations: the ``extended coordinates'' method \cite{Taylor:2022sly,Arrechea:2024cnv,Arrechea:2026mjo} or the ``pragmatic mode-sum'' procedure \cite{Levi:2015eea,Levi:2016quh,Levi:2016paz,Levi:2016exv}.
Recently, the pragmatic mode-sum procedure has been extended \cite{Montagnon:2025vtk} to the computation of the renormalized current for a massless charged scalar field, but this approach has yet to be developed for the RSET. 

In a previous paper \cite{Breen:2024ggu}, employing the extended coordinates approach, we developed an extremely efficient methodology for the computation of {\em {all}} components of both the renormalized current and RSET for a charged, massive, quantum scalar field on a charged, static, spherically-symmetric, black hole background, when the field is in the Hartle-Hawking state \cite{Hartle:1976tp}. 
In this article, we build on this framework to carry out a systematic exploration of the parameter space of both the charged quantum scalar field and the background space-time. We consider the effects on renormalized expectation values of varying the scalar field field mass, charge, and coupling, as well as the charge of the background black hole. 

We restrict our attention in the parameter space to the case where the scalar field is sufficiently massive that charge superradiance is absent \cite{Brito:2015oca,Bekenstein:1973mi,DiMenza:2014vpa,Benone:2015bst}.
 In this situation we can define analogues of  the standard  Hartle-Hawking \cite{Hartle:1976tp}, Unruh \cite{Unruh:1976db} and Boulware \cite{Boulware:1974dm} quantum states. 
 The method in \cite{Breen:2024ggu} was developed using Euclidean techniques and is applicable to a charged quantum scalar field in the Hartle-Hawking state. 
 In this paper, following \cite{Arrechea:2023fas}, we adopt a state subtraction approach using the Hartle-Hawking state as our reference state. 
 In the state subtraction approach one leverages the fact that the difference between expectation values in different quantum states does not require renormalization \cite{Decanini:2005eg}, therefore by calculating $\langle {\hat {J}}^{\alpha } \rangle$ as well as $\langle {\hat {T}}_{\alpha \beta } \rangle$ in the Hartle-Hawking state we obtain their values in the Unruh and Boulware states without recourse to renormalization. 
 This allows us to undertake a comprehensive study of the physical properties of all three states.

This paper is organised as follows. In Sec.~\ref{sec:classical} we construct a basis of mode-solutions to the wave equation satisfied by a charged scalar field in the Reissner-Nordstr{\"o}m black hole space-time. In Sec.~\ref{sec:lorentzian}, we canonically quantize the field and construct the Wightman two-point function for the field in the Unruh, Hartle-Hawking, and Boulware quantum states.  
In Sec. \ref{sec:expvalues} we employ state subtraction to derive expressions for the renormalized current $\langle {\hat {J}}^{\alpha } \rangle$ and RSET $\langle {\hat {T}}_{\alpha \beta}\rangle$ in any quantum state in terms of their values in the 
Hartle-Hawking state. 
We then present and discuss the results of the numerical calculation of these expressions for a wide range of parameters in Sec.~\ref{sec:numerics}. 
Finally we close, in Sec.~\ref{sec:conc}, with our conclusions.

\section{Classical charged scalar field on a charged black hole space-time}
\label{sec:classical}

We consider a scalar field $\Phi $ having mass $\mu >0$ and charge $q$, and with $\xi$ the coupling strength of the field to the background Ricci scalar $R$. The field equation is then given by
\begin{equation}
\left[ D_{\nu }D^{\nu } - \mu ^{2} -\xi\,R \right] \Phi = 0,
    \label{eq:scalar}
\end{equation}
where $D_{\nu }=\nabla _{\nu }-iqA_{\nu }$  is the gauge covariant derivative and $A_{\nu}$ is the electromagnetic gauge potential.
We assume the scalar field propagates on a background Reissner-Nordstr\"om (RN) black hole, whose metric is described by the line element
\begin{equation}
ds^{2} = - f(r) \, dt^{2} + \left[ f(r) \right] ^{-1} dr^{2}+ r^{2} d\theta ^{2} + r^{2}\sin ^{2} \theta \, d\varphi ^{2} ,
\label{eq:RNmetric}
\end{equation}
with metric function $f(r)$ given by 
\begin{equation}
f(r) = 1 - \frac{2M}{r} + \frac{Q^{2}}{r^{2}} ,
\label{eq:fr}
\end{equation}
where $M$ is the mass and $Q$ is the electric charge of the black hole. The electromagnetic gauge potential is $A_{\nu } = (A_{0}, 0 , 0 ,0 )$, with
\begin{equation}
\label{eq:gaugepot}
A_{0} = - \frac {Q}{r} ,
\end{equation}
where we have chosen a gauge in which the electrostatic potential $A_{0}$ tends to zero as $r\rightarrow \infty $, following \cite{Balakumar:2022yvx}.
We restrict attention to the non-extremal black hole solutions where $M^{2}>Q^{2}$,
in which case
the metric function $f(r)$ given by (\ref{eq:fr}) has two zeros, at $r=r_{\pm }$, where
\begin{equation}
r_{\pm } = M \pm {\sqrt {M^{2}-Q^{2}}}.
\label{eq:rpm}
\end{equation}
The larger root of $f(r)$, namely $r_{+}$, is the location of the black hole event horizon 
and $r_{-}$ is the location of the inner Cauchy horizon.
We will be interested only in the region of space-time exterior to the event horizon. 
In this region, we can define the usual ingoing and outgoing null coordinates $u$ and $v$ by
\begin{equation}
    u = t-r_{*}, \qquad v = t+r_{*},
    \label{eq:EF}
\end{equation}
where the ``tortoise'' coordinate $r_{*}$ is defined by 
\begin{equation}
    \frac{dr_{*}}{dr} = \frac{1}{f(r)},
    \label{eq:tortoise}
\end{equation}
and has range $-\infty < r_{*} < \infty $ for $r>r_{+}$. Explicitly, we define
\begin{align}
    r_{*}=r+\frac{r_{+}^{2}}{r_{+}-r_{-}}\ln\left(\frac{r}{r_{+}}-1\right)-\frac{r_{-}^{2}}{r_{+}-r_{-}}\ln\left(\frac{r}{r_{-}}-1\right).
\end{align}
Kruskal coordinates $U$ and $V$ are then given by 
\begin{equation}
    U = -\frac{1}{\kappa }e^{-\kappa u}, \qquad V= \frac{1}{\kappa }e^{\kappa v},
    \label{eq:Kruskal}
\end{equation}
where 
\begin{equation}
    \kappa = \frac{1}{2}f'(r_{+}) = \frac{r_{+}-M}{r_{+}^{2}}
\end{equation}
is the surface gravity of the event horizon.
In the region exterior to the event horizon, we have $U<0$ and $V>0$.

The classical charged scalar field $\Phi $ has a conserved current $J^{\alpha }$ and stress-energy tensor $T_{\alpha \beta }$ given, respectively, by 
\begin{subequations}
\label{eq:classicalobservables}
\begin{equation}
J^{\alpha } = -\frac{q}{4\pi } \Im \left[ \Phi ^{*}D^{\alpha }\Phi \right] ,   
\end{equation}
and 
\begin{align}
T_{\alpha \beta } & = \Re \left\{ \left( 1 - 2\xi \right) \left( D_{\alpha }\Phi \right)^{*} D_{\beta }\Phi
\right. \nonumber \\ & \qquad \left. 
+ \left( 2\xi - \frac{1}{2}\right)  
\frac{1}{2}g_{\alpha \beta }g^{\rho \lambda } \left( D_{\rho }\Phi \right)^{*} D_{\lambda } \Phi 
\right. \nonumber \\ & \qquad \left. 
-2\xi \Phi ^{*} D_{\alpha }D_{\beta } \Phi 
+ 2\xi g_{\alpha \beta } \Phi ^{*} D_{\rho }D^{\rho }\Phi  
\right. \nonumber \\ & \qquad \left. 
+ \xi G_{\mu \nu }  \Phi ^{*}\Phi 
-\frac{1}{2} g_{\alpha \beta } \mu ^{2} \Phi ^{*}\Phi 
\right\} ,
\label{eq:SET}
\end{align}
\end{subequations}
where $\Re$ and $\Im$ denote the real and imaginary parts, respectively, and a superscript ${}^*$ denotes complex conjugation.

Mode solutions of the scalar field equation (\ref{eq:scalar}) take the form
\begin{equation}
    \phi _{\omega \ell m} (t,r,\theta , \varphi ) = \frac{1}{r}{\mathcal {N}}_{\omega }e^{-i\omega t}Y_{\ell m} (\theta , \varphi ) X_{\omega \ell }(r),
    \label{eq:mode}
\end{equation}
where ${\mathcal {N}}_{\omega }$ is a normalization constant, $\omega $ is the mode frequency, $\ell = 0,1, 2, \ldots $ is the total angular momentum quantum number and $m=-\ell , -\ell + 1, \ldots , \ell - 1 ,\ell $ is the azimuthal quantum number.
The functions $Y_{\ell m }(\theta,  \varphi )$ are the usual spherical harmonics, given by 
\begin{equation}
Y_{\ell m }(\theta, \varphi ) = {\sqrt {\frac{(2\ell + 1)}{4\pi }\frac{(\ell - m )!}{(\ell + m )!}}}
P_{\ell }^{m} (\cos \theta ) e^{im\varphi } ,
    \label{eq:spherical}
\end{equation}
where $P_{\ell }^{m}(\cos \theta )$ is an associated Legendre function.
As a result of the definition (\ref{eq:spherical}),
the spherical harmonics are normalized such that
\begin{equation}
\int_{0}^{2\pi }\int _{0}^{\pi } Y_{\ell m }(\theta , \varphi ) Y_{\ell ' m'}^{*}(\theta , \varphi ) \, \sin \theta \, d\theta \, d\varphi = \delta _{\ell \ell'}\delta _{mm'}.
\label{eq:Ynorm}
\end{equation}
In terms of the tortoise coordinate (\ref{eq:tortoise}), the radial function $X_{\omega \ell }(r)$ satisfies the Schr\"odinger-like equation
\begin{equation}
    \left[  -\frac{d^{2}}{dr_{*}^{2}} +V_{\rm {eff}}(r) \right] X_{\omega \ell }(r) =0,
    \label{eq:radialL}
\end{equation}
where the effective potential $V_{\rm {eff}}(r)$ is
\begin{multline}
    V_{{\rm {eff}}}(r) = \frac{f(r)}{r^{2}} \left[ \ell \left(  \ell + 1 \right) + rf'(r) +r^{2}\mu ^{2} \right] \\
- \left( \omega - \frac{qQ}{r}  \right) ^{2}  .
\label{eq:potential}
\end{multline}
The effective potential $V_{\rm{eff}}$ (\ref{eq:potential}) takes the asymptotic values below at the event horizon $r=r_{+}$ ($r_{*} \rightarrow - \infty $) and as $r,r_{*}\rightarrow \infty $:
\begin{equation}
    V_{{\rm {eff}}} (r) \sim 
    \begin{cases}
     - {\widetilde {\omega }}^{2} , & r_{*}\rightarrow - \infty ,\\
     \mu ^{2} -\omega ^{2} , & r_{*} \rightarrow \infty ,
    \end{cases}
    \label{eq:Vlimits}
\end{equation}
where \cite{Balakumar:2022yvx}
\begin{equation}
    {\widetilde {\omega }}  = \omega -  \frac{qQ}{r_{+}}.
    \label{eq:omegatilde}
\end{equation}

Near the event horizon, as $r\rightarrow r_{+}$ and $r_{*}\rightarrow -\infty $, from (\ref{eq:Vlimits}) the solutions of the radial equation (\ref{eq:radialL}) are the usual plane waves:
\begin{equation}
    X_{\omega \ell }(r) \sim e^{\pm i {\widetilde {\omega }} r_{*}}, \qquad {\mbox {as $r_{*}\rightarrow -\infty $}}.
\end{equation}
At infinity, as $r,r_{*} \rightarrow \infty $, the form of the radial function $X_{\omega \ell }(r)$ depends on the sign of $\omega ^{2}-\mu ^{2}$:
\begin{equation}
    X_{\omega \ell }(r) \sim 
    \begin{cases}
        e^{\pm i\Omega  r_{*}}, & \omega ^{2}>\mu ^{2}, 
        \\
        e^{\pm {\widetilde {\Omega }} r_{*}}, & \omega ^{2}<\mu ^{2} ,
    \end{cases}
    \label{eq:infinity}
\end{equation}
where
\begin{align}
        \Omega & =  {\mathrm {sgn}}({\omega }) {\sqrt { \omega ^{2}-\mu ^{2}}} , \qquad \omega ^{2}>\mu ^{2}, 
        \nonumber
        \\
        {\widetilde {\Omega }} & = {\sqrt {\mu ^{2} -  \omega ^{2}}},
        \qquad  \omega ^{2}<\mu ^{2}. 
\end{align}
We note that ${\widetilde {\Omega }}$ is positive for all $-\mu <\omega < \mu $, and the sign of $\Omega $ is the same as the sign of $\omega $.

When $\omega ^{2}<\mu ^{2}$, modes having the positive sign in (\ref{eq:infinity}) are exponentially growing and hence do not correspond to normalizable eigenfunctions for the eigenvalue problem defined by the wave equation (\ref{eq:scalar}). In other words, the interval $\omega^{2}<\mu^{2}$ is not in the spectrum of the differential operator in (\ref{eq:scalar}) with boundary conditions $e^{\tilde{\Omega}r_{*}}$ as $r_{*}\to\infty$. On the other hand, modes having the negative sign in (\ref{eq:infinity}) are exponentially damped and satisfy the boundary conditions for normalizable eigenfunctions. 
In this case, for $\omega^{2}<\mu^{2}$, we define the ``up'' modes as solutions to (\ref{eq:radialL}) with the boundary conditions
\begin{equation}
    X^{\rm {up}}_{\omega \ell}(r) \sim
    \begin{cases}
    e^{i{\widetilde {\omega  }}r_{*}} + A^{\rm {up}}_{\omega \ell }e^{-i{\widetilde {\omega }} r_{*}} , & r_{*}\rightarrow - \infty ,
        \\
       B^{\rm {up}}_{\omega \ell }e^{-{\widetilde {\Omega }}r_{*}} , & r_{*}\rightarrow \infty ,
    \end{cases}
    \label{eq:bound}
\end{equation}
where $A^{\rm {up}}_{\omega \ell }$, $B^{\rm {up}}_{\omega \ell }$ are complex constants, known respectively as the reflection and transmission coefficients for the ``up'' modes.
Although we have denoted these modes as ``up'' modes, since their behaviour near the horizon is the same as the usual ``up'' modes, the behaviour at infinity is different.
The modes (\ref{eq:bound}) are bound state modes. 
As mentioned above, they are normalizable eigenfunctions and hence will need to be included in the two-point function when we consider a quantum scalar field in Sec.~\ref{sec:lorentzian}.

When $\omega ^{2}>\mu ^{2}$, we can define the usual basis of ``in'' and ``up'' modes as solutions to (\ref{eq:radialL}) with boundary conditions
\begin{equation}
    X^{\rm {in}}_{\omega \ell }(r)\sim 
    \begin{cases}
        B^{\rm {in}}_{\omega \ell } e^{-i{\widetilde {\omega }}r_{*}}, & r_{*}\rightarrow - \infty ,
        \\
        e^{-i\Omega r_{*}} + A^{\rm {in}}_{\omega \ell }e^{i\Omega r_{*}},
        & r_{*}\rightarrow \infty ,
    \end{cases}
\end{equation}
and
\begin{equation}
    X^{\rm {up}}_{\omega \ell }(r) \sim 
    \begin{cases}
        e^{i{\widetilde {\omega }}r_{*}} + A^{\rm {up}}_{\omega \ell } e^{-i{\widetilde {\omega }}r_{*}}, & r_{*}\rightarrow - \infty ,
        \\
        B^{\rm {up}}_{\omega \ell }e^{i\Omega r_{*}},
        & r_{*}\rightarrow \infty ,
    \end{cases}
\end{equation}
respectively.
Since the Wronskian of any two linearly independent solutions of the radial equation (\ref{eq:radialL}) is a nonzero constant, the reflection coefficients $A_{\omega \ell }^{\rm {in/up}}$ and transmission coefficients $B_{\omega \ell }^{\rm {in/up}}$ can be shown to satisfy the relations
\begin{subequations}
\label{eq:Wronskian}
\begin{align}
    \Omega  \left[ 1 - \left| A^{\rm {in}}_{\omega \ell} \right| ^{2} \right] & = 
    {\widetilde {\omega }}\left|  B^{\rm {in}}_{\omega \ell} \right| ^{2},
   \\
    {\widetilde {\omega }}\left[ 1 - \left| A^{\rm {up}}_{\omega \ell} \right| ^{2} \right] & = 
    \Omega \left|  B^{\rm {up}}_{\omega \ell} \right| ^{2},
    \\
    {\widetilde {\omega }} B^{\mathrm {in}}_{\omega \ell } & = \Omega B^{\mathrm {up}}_{\omega \ell },
\end{align}
\end{subequations}
for $\omega ^{2}> \mu ^{2}$, while, for $\omega ^{2}<\mu ^{2}$ we have
\begin{equation}
   \left| A^{\rm {up}}_{\omega \ell} \right| ^{2}  = 1.
   \label{eq:BWronskian}
\end{equation}
When $\omega ^{2}<\mu ^{2}$, since the modes (\ref{eq:bound}) are exponentially vanishing at infinity, the amplitudes of the outgoing and ingoing waves near the horizon are the same. 

From (\ref{eq:Wronskian}), it can be seen that if ${\widetilde {\omega }}\Omega <0$, then $| A^{\rm {in/up}}_{\omega \ell}| ^{2}>1$, which means that classical superradiance occurs \cite{Brito:2015oca,Bekenstein:1973mi,DiMenza:2014vpa,Benone:2015bst}.
However, the relations (\ref{eq:Wronskian}) are valid only when $\omega ^{2}>\mu ^{2}$.
Suppose first that $\omega $, $\Omega >0$, then,
in order for superradiant modes to exist, it must be the case that
\begin{equation}
{\widetilde {\omega }} = \omega  - \frac{qQ}{r_{+}} <0.
\label{eq:SRcond1}
\end{equation}
Since $\omega >\mu >0$, this gives
\begin{equation}
    \mu < \frac{qQ}{r_{+}}.
    \label{eq:massbound1}
\end{equation}
Clearly (\ref{eq:massbound1}) can only be satisfied if $qQ>0$ and the scalar field charge has the same sign as the black hole charge.
Now suppose that 
$\omega $, $\Omega <0$, then,
in order for superradiant modes to exist, it must be the case that
\begin{equation}
{\widetilde {\omega }} = \omega  - \frac{qQ}{r_{+}} >0,
\label{eq:SRcond2}
\end{equation}
which gives, since $\omega <-\mu < 0 $,
\begin{equation}
    \mu <- \frac{qQ}{r_{+}}.
    \label{eq:massbound2}
\end{equation}
Eq.~(\ref{eq:massbound2}) can only be satisfied if $qQ<0$ and the black hole and scalar field charges have opposite signs. 
To summarize, classical superradiant modes do not exist if
\begin{equation}
    \mu > \frac{|qQ|}{r_{+}} ,
    \label{eq:SRbound}
\end{equation}
and we shall assume for the rest of this paper that (\ref{eq:SRbound}) holds (see \cite{DiMenza:2014vpa} for a related result on the absence of superradiance if the scalar field mass $\mu $ satisfies (\ref{eq:SRbound})). 

The scalar field modes are normalized using the inner product
\begin{equation}
    \langle \Phi _{1} , \Phi _{2}   \rangle
    = i\int _{\Sigma } \left[ \left(  D_{\nu }\Phi _{1} \right) ^{*} \Phi _{2} -  \Phi _{1}^{*} D_{\nu } \Phi _{2} \right]  {\sqrt {-g}} \, d\Sigma ^{\nu },
\end{equation}
where the integral is performed over a Cauchy surface $\Sigma $.
We find
\begin{align}
 \langle \phi ^{\rm {in}}_{\omega \ell m } , \phi ^{\rm {in}}_{\omega '\ell ' m'}   \rangle
 & = 
 4\pi \Omega {\mathcal {N}}_{\omega }^{\rm {in}}{}^{*}{\mathcal {N}}_{\omega '}^{\rm {in}}
 \delta (\omega  -\omega ') \delta _{\ell \ell '}\delta _{m m'},
 \nonumber \\
  \langle \phi ^{\rm {up}}_{\omega \ell m } , \phi ^{\rm {up}}_{\omega '\ell ' m'}   \rangle
 & = 
 4\pi {\widetilde {\omega  }}{\mathcal {N}}_{\omega }^{\rm {up}}{}^{*}{\mathcal {N}}_{\omega '}^{\rm {up}}
 \delta (\omega  -\omega ') \delta _{\ell \ell '}\delta _{m m'},
\end{align}
so that we can take the normalization constants to be
\begin{equation}
    {\mathcal {N}}^{\rm {in}}_{\omega } = \frac{1}{{\sqrt { 4\pi |\Omega |}}}, \qquad
    {\mathcal {N}}^{\rm {up}}_{\omega } = \frac{1}{{\sqrt {4\pi |{\widetilde {\omega }} |}}}.
\end{equation}
The ``in'' modes have positive ``norm'' if $\Omega >0$, while the ``up'' modes have positive ``norm'' if ${\widetilde {\omega }}>0$. 
Our earlier analysis shows that, when the scalar field mass $\mu $ satisfies (\ref{eq:SRbound}), then $\Omega $ and ${\widetilde {\omega }}$ have the same sign.
This will greatly simplify the quantization of the scalar field, to which we now proceed.

\section{Quantum charged scalar field on a charged black hole space-time}
\label{sec:lorentzian}

We now use canonical quantization to define the usual quantum states (Hartle-Hawking \cite{Hartle:1976tp}, Unruh \cite{Unruh:1976db} and Boulware \cite{Boulware:1974dm}) for the charged scalar field on RN, assuming that the mass bound (\ref{eq:SRbound}) holds.
The absence of superradiant modes in this case means that the construction of states is much simpler than that in \cite{Balakumar:2022yvx} for the massless charged scalar.

\subsection{Boulware state}
\label{sec:Boulware}

To construct the Boulware state $|{\rm {B}}\rangle $, we define positive $\phi ^{{\rm {in/up}}+}_{\omega \ell m}$ and negative $\phi ^{{\rm {in/up}}-}_{\omega \ell m}$ frequency modes as follows:
\begin{align}
    \phi ^{{\rm {in}}+}_{\omega \ell m} & = \frac{e^{-i\omega t}}{{\sqrt {4\pi |\Omega |}}r} X^{\rm {in}}_{\omega \ell }(r) Y_{\ell m}(\theta , \varphi ), \qquad \omega >\mu ,
    \nonumber \\
    \phi ^{{\rm {in}}-}_{\omega \ell m} & = \frac{e^{-i\omega t}}{{\sqrt {4\pi |\Omega |}}r} X^{\rm {in}}_{\omega \ell }(r) Y_{\ell m}(\theta , \varphi ), \qquad \omega < -\mu ,
    \nonumber \\
    \phi ^{{\rm {up}}+}_{\omega \ell m} & = \frac{e^{-i\omega t}}{{\sqrt {4\pi |{\widetilde {\omega }}|}}r} X^{\rm {up}}_{\omega \ell }(r) Y_{\ell m}(\theta , \varphi ), \qquad {\widetilde {\omega }}>0 ,
    \nonumber \\
    \phi ^{{\rm {up}}-}_{\omega \ell m} & = \frac{e^{-i\omega t}}{{\sqrt {4\pi |{\widetilde {\omega }}|}}r} X^{\rm {up}}_{\omega \ell }(r) Y_{\ell m}(\theta , \varphi ), \qquad {\widetilde {\omega }}<0,
\label{eq:Bmodes}
\end{align}
We expand the scalar field $\Phi $ in terms of the above orthonormal basis of field modes, and then quantize the field by promoting the expansion coefficients to operators, giving the quantum scalar field ${\hat {\Phi }}$ to be
\begin{align}
    {\hat {\Phi }} & = 
    \sum _{\ell =0}^{\infty } \sum _{m=-\ell }^{\ell } \left\{ 
    \int _{\mu }^{\infty } d\omega \, 
    {\hat {a}}^{\rm {in}}_{\omega \ell m} \phi ^{{\rm {in}}+}_{\omega \ell m}
    + \int _{-\infty }^{-\mu } d\omega \, 
    {\hat {b}}^{{\rm {in}}\dagger }_{\omega \ell m} \phi ^{{\rm {in}}-}_{\omega \ell m}
    \right.
    \nonumber \\ &  \qquad 
    \left. 
    +\int _{0}^{\infty } d{\widetilde {\omega }}\, 
    {\hat {a}}^{\rm {up}}_{\omega \ell m} \phi ^{{\rm {up}}+}_{\omega \ell m}
    + \int _{-\infty }^{0} d{\widetilde {\omega }}\, 
    {\hat {b}}^{{\rm {up}}\dagger }_{\omega \ell m} \phi ^{{\rm {up}}-}_{\omega \ell m} 
    \right\} .
\end{align}
The operators ${\hat {a}}^{\rm {in}}_{\omega \ell m}$, ${\hat {a}}^{\rm {up}}_{\omega \ell m}$,
${\hat {b}}^{\rm {in}}_{\omega \ell m}$ and ${\hat {b}}^{\rm {up}}_{\omega \ell m}$
satisfy the usual commutation relations
\begin{align}
    \left[  {\hat {a}}^{\rm {in}}_{\omega \ell m}, {\hat {a}}^{{\rm {in}}\dagger }_{\omega '\ell 'm'} \right] & = \delta (\omega -\omega ') \delta _{\ell \ell '}\delta _{m m'},
    \qquad \omega >\mu ,
    \nonumber \\ 
     \left[  {\hat {b}}^{\rm {in}}_{\omega \ell m}, {\hat {b}}^{{\rm {in}}\dagger }_{\omega '\ell 'm'} \right] & = \delta (\omega -\omega ') \delta _{\ell \ell '}\delta _{m m'},
    \qquad \omega <-\mu ,
    \nonumber \\
     \left[  {\hat {a}}^{\rm {up}}_{\omega \ell m}, {\hat {a}}^{{\rm {up}}\dagger }_{\omega '\ell 'm'} \right] & = \delta (\omega -\omega ') \delta _{\ell \ell '}\delta _{m m'},
    \qquad {\widetilde {\omega }}>0,
    \nonumber \\ 
     \left[  {\hat {b}}^{\rm {up}}_{\omega \ell m}, {\hat {b}}^{{\rm {up}}\dagger }_{\omega '\ell 'm'} \right] & = \delta (\omega -\omega ') \delta _{\ell \ell '}\delta _{m m'},
    \qquad {\widetilde {\omega }}<0,
\end{align}
with all other commutators vanishing.
The Boulware state $|{\rm {B}}\rangle $ is then defined as the state annihilated by the operators ${\hat {a}}^{\rm {in}}_{\omega \ell m}$, ${\hat {a}}^{\rm {up}}_{\omega \ell m}$,
${\hat {b}}^{\rm {in}}_{\omega \ell m}$ and ${\hat {b}}^{\rm {up}}_{\omega \ell m}$:
\begin{align}
    {\hat {a}}^{\rm {in}}_{\omega \ell m} |{\rm {B}}\rangle   & = 0, \qquad \omega > \mu ,
    \nonumber \\
    {\hat {b}}^{\rm {in}}_{\omega \ell m} |{\rm {B}}\rangle   & = 0, \qquad \omega < -\mu ,
    \nonumber \\
    {\hat {a}}^{\rm {up}}_{\omega \ell m} |{\rm {B}}\rangle   & = 0, \qquad {\widetilde {\omega }} >0,
    \nonumber \\
    {\hat {b}}^{\rm {up}}_{\omega \ell m} |{\rm {B}}\rangle   & = 0, \qquad {\widetilde {\omega  }}<0.
\end{align}
In this construction, we have not had to distinguish between a ``past'' and ``future'' Boulware state, unlike the situation for a massless charged scalar field \cite{Balakumar:2022yvx}. 
While we have defined the state $|{\mathrm {B}}\rangle $ using the ``in'' and ``up'' basis modes, we could equally well have considered a basis consisting of the time-reverse of these modes (namely the ``out'' and ``down'' modes \cite{Balakumar:2022yvx}). 
Using the ``out'' and ``down'' modes as a basis in the above construction would yield exactly the same state $|{\mathrm {B}}\rangle $, since we have no superradiant modes.

The (unrenormalized) expectation value of an operator ${\hat {O}}$ (corresponding to a classical quantity $O$) when the charged scalar field is in the Boulware state $|{\mathrm {B}}\rangle $ is given by 
\begin{multline}
    \langle {\rm {B}} | {\hat {O}} | {\rm {B}} \rangle 
    = 
    \frac{1}{2} \sum _{\ell  =0}^{\infty } \sum _{m=-\ell }^{\ell }  \left\{ 
     \int _{|\omega | > \mu } d\omega \, o_{\omega \ell m}^{\rm {in}} 
     \right.  \\ \left. 
     + \int _{-\infty }^{\infty } d{\widetilde {\omega }}
    \, o_{\omega \ell m}^{\rm {up}}   \right\} ,
    \label{eq:Bexp}
\end{multline}
where $o_{\omega \ell m}^{\rm {in/up}}$ is the classical value of $O$ computed for a particular in/up mode. 
For example, if ${\hat {O}}$ is the scalar condensate ${\hat {\Phi }}^{2}$, then $O$ is simply the square of the absolute value of the classical scalar field and $o_{\omega \ell m}^{\rm {in/up}}= |\phi _{\omega \ell m }|^{2}$.

In this paper, as well as the scalar condensate, we are particularly interested in the renormalized expectation values of the charged scalar current operator $\langle {\hat {J}}^{\alpha }\rangle $ and the stress-energy tensor operator $\langle {\hat {T}}_{\alpha \beta }\rangle $ [whose corresponding classical quantities are given in (\ref{eq:classicalobservables})].
Like the classical charge current, the expectation value $\langle {\hat {J}}^{\alpha }\rangle $ is conserved \cite{Balakumar:2019djw}:
\begin{equation}
\label{eq:currentcons}
    \nabla _{\alpha } \langle {\hat {J}} ^{\alpha } \rangle = 0,
\end{equation}
which, for time-independent states on the static RN background, reduces to a simple equation which 
is readily integrated to give the following expression for the radial component of the renormalized charged scalar current \cite{Balakumar:2022yvx}:
\begin{equation}
\label{eq:Jr}
    \langle {\hat {J}} ^{r} \rangle  = -\frac {\mathcal {K}}{r^{2}},
\end{equation}
where ${\mathcal {K}}$ is a state-dependent constant representing the flux of charge emitted by the black hole.

The coupling between the quantum scalar field and the background electromagnetic field means that the expectation value $\langle {\hat {T}}_{\alpha \beta }\rangle $ is not conserved, but instead satisfies the equation \cite{Balakumar:2019djw}:
\begin{equation}
\label{eq:conservation}
    \nabla ^{\alpha }\langle {\hat {T}}_{\alpha \beta } \rangle = 4\pi F_{\alpha \beta } \langle {\hat {J}}^{\alpha } \rangle .
\end{equation}
For static states, setting $\beta = t$ in (\ref{eq:conservation})  again gives a simple equation whose integration is straightforward, yielding the solution
\begin{equation}
\label{eq:Trt}
\langle {\hat {T}}^{r}{}_{t} \rangle = -\frac{\mathcal {L}}{r^{2}} + \frac{4\pi Q{\mathcal {K}}}{r^{3}},
\end{equation}
where ${\mathcal {L}}$ is a second state-dependent constant, corresponding to the flux of energy emitted by the black hole. 
Since neither $ \langle {\hat {J}} ^{r} \rangle$ nor $\langle {\hat {T}}_{t}^{r} \rangle $ require renormalization \cite{Balakumar:2022yvx}, it is straightforward to compute the fluxes ${\mathcal {K}}$ and ${\mathcal {L}}$ for any quantum state of interest. 
 
In the Boulware state $|{\mathrm {B}}\rangle $, we find that the fluxes of charge ${\mathcal {K}}_{{\mathrm {B}}}$ and energy ${\mathcal {L}}_{{\mathrm {B}}}$ both vanish. 
This is in contrast to the situation for a massless charged scalar field \cite{Balakumar:2020gli, Balakumar:2022yvx} where the ``past'' Boulware state $|{\mathrm {B}}^{-}\rangle $ contains an outgoing flux of both charge and energy from the black holes, contributed entirely by the superradiant modes.
Since we are considering here a charged scalar field whose mass is sufficiently large that superradiance is absent, there is no flux of charge or energy in the Boulware state. 
The form of the unrenormalized expectation value (\ref{eq:Bexp}) suggests that the Boulware state corresponds to an absence of particles in either the ``in'' or ``up'' modes, and therefore we expect that this state will be as empty as possible as seen by a static observer far from the black hole.

\subsection{Unruh state}
\label{sec:Unruh}

To define the Unruh state $|{\rm {U}}\rangle $, we employ the positive and negative frequency ``in'' modes as in (\ref{eq:Bmodes}).
For the ``up'' modes, we follow the procedure in \cite{Balakumar:2022yvx} and define the following set of modes, which are, respectively, positive and negative frequency with respect to Kruskal time on a surface close to the past horizon of the RN space-time:
\begin{align}
    \chi ^{{\rm {up}}+}_{\omega \ell m } & =
    \frac{1}{{\sqrt {2 \sinh \left| \frac{\pi {\widetilde {\omega }}}{\kappa } \right|}}} \left(
    e^{\frac{\pi {\widetilde {\omega }}}{2\kappa }}\phi ^{\rm {up}}_{\omega \ell m} 
    + e^{-\frac{\pi {\widetilde {\omega }}}{2\kappa }}\psi  ^{\rm {down}}_{\omega \ell m}
    \right) ,
    \quad \forall \, {\widetilde {\omega }} ,
    \nonumber \\
    \chi ^{{\rm {up}}-}_{\omega \ell m } & =
    \frac{1}{{\sqrt {2 \sinh \left| \frac{\pi {\widetilde {\omega }}}{\kappa } \right|}}} \left(
    e^{-\frac{\pi {\widetilde {\omega }}}{2\kappa }}\phi ^{\rm {up}}_{\omega \ell m} 
    + e^{\frac{\pi {\widetilde {\omega }}}{2\kappa }}\psi  ^{\rm {down}}_{\omega \ell m}
    \right) ,
    \quad \forall \, {\widetilde {\omega }}.
    \label{eq:Umodes}
\end{align}
The $\phi ^{\rm {up}}_{\omega \ell m}$ modes are defined for {\em {all}} frequencies by
\begin{equation}
     \phi ^{{\rm {up}}}_{\omega \ell m} = \frac{e^{-i\omega t}}{{\sqrt {4\pi |{\widetilde {\omega }}|}}r} X^{\rm {up}}_{\omega \ell }(r) Y_{\ell m}(\theta , \varphi ), \qquad \forall \, {\widetilde {\omega }} .
\end{equation}
The modes $\psi ^{\rm {down}}_{\omega \ell m}$ are defined by making the transformation $(U,V)\rightarrow (-U,-V)$ of the Kruskal coordinates (\ref{eq:Kruskal}), and are nonzero only on the left-hand-diamond of the RN conformal diagram.
Expanding the quantum scalar field in terms of the modes $\phi ^{{\rm {in}}+}_{\omega \ell m}$, $\phi ^{{\rm {in}}-}_{\omega \ell m}$ (\ref{eq:Bmodes}) and $\chi ^{{\rm {up}}+}_{\omega \ell m } $,  $\chi ^{{\rm {up}}-}_{\omega \ell m } $ (\ref{eq:Umodes}) gives
\begin{widetext}
\begin{align}
     {\hat {\Phi }} & = 
    \sum _{\ell =0}^{\infty } \sum _{m=-\ell }^{\ell } \left\{ 
    \int _{\mu }^{\infty } d\omega \, 
    {\hat {c}}^{\rm {in}}_{\omega \ell m} \phi ^{{\rm {in}}+}_{\omega \ell m}
    + \int _{-\infty }^{-\mu } d\omega \, 
    {\hat {d}}^{{\rm {in}}\dagger }_{\omega \ell m} \phi ^{{\rm {in}}-}_{\omega \ell m}
    +
    \int _{-\infty }^{\infty }d{\widetilde {\omega }} \, \frac{\phi ^{\rm {up}}_{\omega \ell m}}{{\sqrt {2 \sinh \left| \frac{\pi {\widetilde {\omega }} }{\kappa } \right|}}} 
    \left[ 
    e^{\frac{\pi {\widetilde {\omega }} }{2\kappa }} {\hat {c}}^{\rm {up}}_{\omega \ell m}  +
    e^{-\frac{\pi {\widetilde {\omega }} }{2\kappa }} {\hat {d}}^{{\rm {up}}\dagger }_{\omega \ell m} 
    \right] 
    \right\} ,
\end{align}
\end{widetext}
where we have restricted our attention to the right-hand-wedge of the RN conformal diagram, on which the modes $\psi ^{\rm {down}}_{\omega \ell m}$ vanish.

The nonvanishing commutation relations satisfied by the  operators ${\hat {c}}^{\rm {in}}_{\omega \ell m}$, ${\hat {c}}^{\rm {up}}_{\omega \ell m}$,
${\hat {d}}^{\rm {in}}_{\omega \ell m}$ and ${\hat {d}}^{\rm {up}}_{\omega \ell m}$
are
\begin{align}
    \left[  {\hat {c}}^{\rm {in}}_{\omega \ell m}, {\hat {c}}^{{\rm {in}}\dagger }_{\omega '\ell 'm'} \right] & = \delta (\omega -\omega ') \delta _{\ell \ell '}\delta _{m m'},
    \qquad \omega > \mu  ,
    \nonumber \\ 
     \left[  {\hat {d}}^{\rm {in}}_{\omega \ell m}, {\hat {d}}^{{\rm {in}}\dagger }_{\omega '\ell 'm'} \right] & = \delta (\omega -\omega ') \delta _{\ell \ell '}\delta _{m m'},
    \qquad \omega < -\mu ,
    \nonumber \\
     \left[  {\hat {c}}^{\rm {up}}_{\omega \ell m}, {\hat {c}}^{{\rm {up}}\dagger }_{\omega '\ell 'm'} \right] & = \delta (\omega -\omega ') \delta _{\ell \ell '}\delta _{m m'},
    \qquad \forall \, {\widetilde {\omega }} ,
    \nonumber \\ 
     \left[  {\hat {d}}^{\rm {up}}_{\omega \ell m}, {\hat {d}}^{{\rm {up}}\dagger }_{\omega '\ell 'm'} \right] & = \delta (\omega -\omega ') \delta _{\ell \ell '}\delta _{m m'},
    \qquad \forall \, {\widetilde {\omega }} .
\end{align}
The Unruh state $|{\rm {U}}\rangle $ is then defined as follows:
\begin{align}
    {\hat {c}}^{\rm {in}}_{\omega \ell m} |{\rm {U}}\rangle  & = 0, \qquad \omega > \mu ,
    \nonumber \\
    {\hat {d}}^{\rm {in}}_{\omega \ell m} |{\rm {U}}\rangle  & = 0, \qquad \omega < - \mu ,
    \nonumber \\
    {\hat {c}}^{\rm {up}}_{\omega \ell m} |{\rm {U}}\rangle  & = 0, \qquad \forall \, {\widetilde {\omega }} ,
    \nonumber \\
    {\hat {d}}^{\rm {up}}_{\omega \ell m} |{\rm {U}}\rangle   & = 0, \qquad  \forall  \, {\widetilde {\omega }} .
\end{align}
The state $|{\mathrm {U}}\rangle $ defined here corresponds to the ``past'' Unruh state constructed in Ref.~\cite{Balakumar:2022yvx}, and is the most relevant state modelling the emission of Hawking radiation by a black hole formed from gravitational collapse.
It would be possible, following \cite{Balakumar:2022yvx}, to construct a ``future'' Unruh state, but the physical relevance of such a state is not clear and we do not consider it further. 

The (unrenormalized) expectation value of an operator ${\hat {O}}$ when the charged scalar field is in the Unruh state $|{\mathrm {U}}\rangle $ is given by 
\begin{multline}
    \langle {\rm {U}} | {\hat {O}} | {\rm {U}} \rangle 
     = 
    \frac{1}{2} \sum _{\ell  =0}^{\infty } \sum _{m=-\ell }^{\ell } 
    \left\{ 
     \int _{|\omega | > \mu } d\omega
    \,  o_{\omega \ell m}^{\rm {in}}
    \right. \nonumber \\  \left. 
    +   \int _{-\infty }^{\infty } d{\widetilde {\omega }} \, o_{\omega \ell m}^{\rm {up}}  \coth \left|  \frac{\pi {\widetilde {\omega }}}{\kappa } \right|  \right\}  .
    \label{eq:Uexp}
\end{multline}
Unlike the Boulware state $|{\mathrm {B}}\rangle $, the Unruh state $|{\mathrm {U}}\rangle $ is not an equilibrium state, as it has nonzero fluxes of charge ${\mathcal {K}}_{{\mathrm {U}}}$ and energy ${\mathcal{L}}_{{\mathrm {U}}}$ \cite{Balakumar:2022yvx}:
\begin{subequations}
\label{eq:Ufluxes}
\begin{align}
    {\mathcal {K}}_{{\rm {U}}} & = \frac{q}{64\pi ^{3}} \sum _{\ell =0}^{\infty } 
    \int _{|\omega | > \mu } d\omega \, \left( 2 \ell + 1 \right) \frac{\Omega }{|{\widetilde {\omega }}|}
    \frac{1}{e^{\frac{2\pi |{\widetilde {\omega }} |}{\kappa }}-1} | B^{{\rm {up}}}_{\omega \ell } |^{2} ,
    \\
    {\mathcal {L}}_{\rm {U}} & = \frac{1}{16\pi ^{2}}
    \sum _{\ell =0}^{\infty } 
    \int _{|\omega | > \mu } d\omega  \, \left( 2 \ell + 1 \right) \frac{\Omega  \omega }{|{\widetilde {\omega }} |}
    \frac{1}{e^{\frac{2\pi |{\widetilde {\omega }} |}{\kappa }}-1} | B^{{\rm {up}}}_{\omega \ell } |^{2} .
\end{align}
\end{subequations}
These expressions take a slightly different form to those in \cite{Balakumar:2022yvx} due to the nonzero mass of the charged scalar field.
The expressions (\ref{eq:Ufluxes}) reveal thermal emission of charged scalar particles in the ``up'' modes, with an effective chemical potential \cite{Gibbons:1975kk,Hawking:1975vcx}. 
Since, by definition, we have $\omega \Omega >0$, the flux of energy ${\mathcal {L}}_{\mathrm {U}}$ from the black hole is always positive, as expected. 
In contrast, the flux of charge ${\mathcal {K}}_{\mathrm {U}}$ may have either sign, depending on the details of the relative contribution from the positive and negative frequency modes and the sign of the scalar field charge $q$.

\subsection{Hartle-Hawking state}
\label{sec:HH}

The final state we consider is the Hartle-Hawking state $|{\rm {H}}\rangle $. 
We now employ the modes $\chi ^{{\rm {up}}+}_{\omega \ell m } $, $\chi ^{{\rm {up}}-}_{\omega \ell m  }$ (\ref{eq:Umodes}) as for the construction of the Unruh state, but instead of the ``in'' modes (\ref{eq:Bmodes}) we will use the following mode functions, defined for all $\omega $ such that $|\omega |>\mu $:
\begin{align}
    \chi ^{{\rm {in}}+}_{\omega \ell m } & =
    \frac{1}{{\sqrt {2 \sinh \left| \frac{\pi {\widetilde {\omega }}}{\kappa } \right|}}} \left(
    e^{\frac{\pi {\widetilde {\omega }}}{2\kappa }}\phi ^{\rm {in}}_{\omega \ell m} 
    + e^{-\frac{\pi {\widetilde {\omega }} }{2\kappa }}\psi  ^{\rm {out}}_{\omega \ell m}
    \right) ,
    \nonumber \\
    \chi ^{{\rm {in}}-}_{\omega \ell m } & =
    \frac{1}{{\sqrt {2 \sinh \left| \frac{\pi {\widetilde {\omega }}}{\kappa } \right|}}} \left(
    e^{-\frac{\pi {\widetilde {\omega }} }{2\kappa }}\phi ^{\rm {in}}_{\omega \ell m} 
    + e^{\frac{\pi {\widetilde {\omega }}}{2\kappa }}\psi  ^{\rm {out}}_{\omega \ell m}
    \right)  .
    \label{eq:Hmodes}
\end{align}
We now have 
\begin{equation}
    \phi ^{{\rm {in}}}_{\omega \ell m}  = \frac{e^{-i\omega t}}{{\sqrt {4\pi |\Omega |}}r} X^{\rm {in}}_{\omega \ell }(r) Y_{\ell m}(\theta , \varphi ), \qquad \forall \, |\omega |>\mu ,
\end{equation}
and define the $\psi  ^{\rm {out}}_{\omega \ell m}$ modes from the $\phi  ^{\rm {in}}_{\omega \ell m}$ modes by applying the change of coordinates $(U,V)\rightarrow (-U,-V)$.

Following the analysis in \cite{Balakumar:2022yvx}, the modes (\ref{eq:Hmodes}) have positive frequency with respect to Kruskal time on a surface close to the future horizon of the black hole. 
The quantum scalar field is expanded in terms of the modes (\ref{eq:Umodes}, \ref{eq:Hmodes}) as follows (on the right-hand-diamond of the RN Kruskal diagram):
\begin{widetext}
\begin{align}
     {\hat {\Phi }} & = 
    \sum _{\ell =0}^{\infty } \sum _{m=-\ell }^{\ell } \left\{ 
    \int _{|\omega |>\mu } d\omega \, 
    \frac{\phi ^{\rm {in}}_{\omega \ell m}}{{\sqrt {2 \sinh \left| \frac{\pi {\widetilde {\omega }} }{\kappa } \right|}}} 
    \left[ 
    e^{\frac{\pi {\widetilde {\omega }} }{2\kappa }} {\hat {f}}^{\rm {in}}_{\omega \ell m}  +
    e^{-\frac{\pi {\widetilde {\omega }} }{2\kappa }} {\hat {g}}^{{\rm {in}}\dagger }_{\omega \ell m} 
    \right] 
    +
    \int _{-\infty }^{\infty }d{\widetilde {\omega }} \, 
    \frac{\phi ^{\rm {up}}_{\omega \ell m}}{{\sqrt {2 \sinh \left| \frac{\pi {\widetilde {\omega }} }{\kappa } \right|}}} 
    \left[ 
    e^{\frac{\pi {\widetilde {\omega }} }{2\kappa }} {\hat {f}}^{\rm {up}}_{\omega \ell m}  +
    e^{-\frac{\pi {\widetilde {\omega }} }{2\kappa }} {\hat {g}}^{{\rm {up}}\dagger }_{\omega \ell m} 
    \right] 
    \right\} .
\end{align}
\end{widetext}
The operators ${\hat {f}}^{\rm {in}}_{\omega \ell m}$, ${\hat {f}}^{\rm {up}}_{\omega \ell m}$, 
${\hat {g}}^{\rm {in}}_{\omega \ell m}$ and ${\hat {g}}^{\rm {up}}_{\omega \ell m}$
satisfy the commutation relations (all commutators not given explicitly below vanish):
\begin{align}
    \left[  {\hat {f}}^{\rm {in}}_{\omega \ell m}, {\hat {f}}^{{\rm {in}}\dagger }_{\omega '\ell 'm'} \right] & = \delta (\omega -\omega ') \delta _{\ell \ell '}\delta _{m m'},
    \qquad \forall \, |\omega |>\mu ,
    \nonumber \\ 
     \left[  {\hat {g}}^{\rm {in}}_{\omega \ell m}, {\hat {g}}^{{\rm {in}}\dagger }_{\omega '\ell 'm'} \right] & = \delta (\omega -\omega ') \delta _{\ell \ell '}\delta _{m m'},
    \qquad \forall  \, |\omega |>\mu ,
    \nonumber \\
     \left[  {\hat {f}}^{\rm {up}}_{\omega \ell m}, {\hat {f}}^{{\rm {up}}\dagger }_{\omega '\ell 'm'} \right] & = \delta (\omega -\omega ') \delta _{\ell \ell '}\delta _{m m'},
    \qquad \forall \, {\widetilde {\omega }} ,
    \nonumber \\ 
     \left[  {\hat {g}}^{\rm {up}}_{\omega \ell m}, {\hat {g}}^{{\rm {up}}\dagger }_{\omega '\ell 'm'} \right] & = \delta (\omega -\omega ') \delta _{\ell \ell '}\delta _{m m'},
    \qquad \forall \, {\widetilde {\omega }} .
\end{align}
We then define the Hartle-Hawking state $|{\rm {H}}\rangle $ by:
\begin{align}
    {\hat {f}}^{\rm {in}}_{\omega \ell m} |{\rm {H}}\rangle   & = 0, \qquad \forall \, |\omega |>\mu  ,
    \nonumber \\
    {\hat {g}}^{\rm {in}}_{\omega \ell m} |{\rm {H}}\rangle   & = 0, \qquad \forall \, |\omega |>\mu  ,
    \nonumber \\
    {\hat {f}}^{\rm {up}}_{\omega \ell m} |{\rm {H}}\rangle   & = 0, \qquad \forall \, {\widetilde {\omega }} ,
    \nonumber \\
    {\hat {g}}^{\rm {up}}_{\omega \ell m} |{\rm {H}}\rangle   & = 0, \qquad  \forall  \, {\widetilde {\omega }}.
\end{align}
As for the Boulware state, the absence of superradiant modes has greatly simplified the construction of the state $|{\mathrm {H}}\rangle $ compared to the massless charged scalar field case \cite{Balakumar:2022yvx}.
The state defined here is the analogue of the ``Hartle-Hawking-like'' state (also denoted by $|{\mathrm {H}}\rangle $ in \cite{Balakumar:2022yvx}), whose construction involved a nonstandard treatment of the quantization of the superradiant ``in'' modes. 
There are no such subtleties here, since there are no superradiant modes. 

The (unrenormalized) expectation value of an operator ${\hat {O}}$ when the charged scalar field is in the Hartle-Hawking state is given by 
\begin{multline}
    \langle {\rm {H}} | {\hat {O}} | {\rm {H}} \rangle 
     = 
    \frac{1}{2} \sum _{\ell  =0}^{\infty } \sum _{m=-\ell }^{\ell } 
    \left\{ 
     \int _{|\omega |>\mu } d\omega
   \,  o_{\omega \ell m}^{\rm {in}} \coth \left|  \frac{\pi {\widetilde {\omega }} }{\kappa } \right| 
   \right.  \\ \left.  
   +  \int _{-\infty }^{\infty } d{\widetilde {\omega }}
   \,  o_{\omega \ell m}^{\rm {up}}   \coth \left|  \frac{\pi {\widetilde {\omega }}}{\kappa } \right|  \right\}  .
   \label{eq:Hexp}
\end{multline}
We find that, like the Boulware state $|{\mathrm {B}}\rangle $, the Hartle-Hawking state $|{\mathrm {H}}\rangle $ is an equilibrium state with vanishing fluxes of both charge and energy. 
However, unlike the Boulware state, the form of the unrenormalized expectation value (\ref{eq:Hexp}) implies that the Hartle-Hawking state is a thermal equilibrium state, with a thermal distribution of particles in both the ``in'' and ``up'' modes.

\section{Renormalized expectation values}
\label{sec:expvalues}

Having constructed the quantum states of interest for the charged scalar field, the next step is to outline our procedure for computing  the renormalized expectation values of the physical quantities of interest. 
 
Let the charged quantum scalar field $\hat{\Phi}$ be in the unit norm state $|A\rangle$, then we define the two-point function
\begin{equation}
    G_{A}(x,x')=\langle A|\{\hat{\Phi}(x),\hat{\Phi}^{\dagger}(x')\}|A\rangle,
\end{equation}
where 
\begin{equation}
\{\hat{\Phi}(x),\hat{\Phi}^{\dagger}(x')\}=\frac{1}{2}\left[\hat{\Phi}(x)\hat{\Phi}^{\dagger}(x')+\hat{\Phi}^{\dagger}(x')\hat{\Phi}(x)\right] .
\end{equation}
This two-point function is sesquisymmetric in $x$ and $x'$, that is,
\begin{align}
    G_{A}(x,x')=G_{A}^{\dagger}(x',x).
\end{align}
It is evident that $G_{A}(x,x')$ is ill-defined at the coincidence limit $x'\to x$. Provided the state $|A\rangle$ is a Hadamard state, the divergences in this limit are universal in the sense that they depend only the geometry of the metric (and its derivatives) and the field parameters, and importantly are independent of the quantum state. 
In particular, the Hadamard parametrix for the wave equation (\ref{eq:scalar}) encapsulates the short-distance divergences in $G_{A}(x,x')$ and is given by \cite{Decanini:2005eg,Balakumar:2019djw} 
\begin{align}
\label{eq:KHad}
    K(x,x')=
    \frac{1}{8\pi^{2}}\left\{ \frac{U(x,x')}{\sigma_{\varepsilon}(x,x')}+V(x,x')\log{\left[\frac{2\sigma_{\varepsilon}(x,x')}{L^{2}}\right]}\right\},
\end{align}
where $\sigma_{\varepsilon}(x,x')\equiv\sigma(x,x')+i\varepsilon$, with $\sigma(x,x')$ being Synge's world function (corresponding to half the square of the geodesic distance between the points $x$ and $x'$). 
The $i\varepsilon$ is required to define the parametrix as a distribution since $\sigma(x,x')$ vanishes not only at coincidence $x=x'$ but also when $x$ and $x'$ are connected by a null geodesic.

The biscalars $U(x,x')$ and $V(x,x')$ are sesquisymmetric in the space-time points and are regular in the coincidence limit. 
It is essential to the renormalizability of the semiclassical theory that these biscalars are constructed locally from the metric and its derivatives \cite{Balakumar:2019djw}. 
An arbitrary length scale $L$ has been inserted into the $\log $ term  in (\ref{eq:KHad}) to make the argument dimensionless. 
This arbitrariness is part of the well-known renormalization ambiguity \cite{Decanini:2005eg,Balakumar:2019djw} and is a manifestation of the fact that the biscalar $V(x,x')$ is a solution of the charged scalar field homogeneous wave equation \cite{Balakumar:2019djw}
\begin{equation}
\label{eq:Veqn}
    \left[ D_{\alpha }D^{\alpha }-(\mu^{2}+\xi R)\right]  V(x,x')=0,
\end{equation}
and so we are free to add multiples of $V(x,x')$ to any parametrix. 
The biscalar $U(x,x')$ satisfies the transport equation  \cite{Balakumar:2019djw} 
\begin{subequations}
    \label{eq:Ueqn}
\begin{align}
\left[ 2\sigma^{;\alpha }D_{\alpha }+\Box\sigma-4\right] & U(x,x')=0,
\end{align}
subject to the boundary condition
\begin{equation}
    U(x,x)=1.
\end{equation}
\end{subequations}
Eq.~(\ref{eq:Ueqn}) is readily solved as a covariant Taylor series about one of the points. 

We solve (\ref{eq:Veqn}) by expanding
\begin{equation}
\label{eq:Vdef}
    V(x,x') = \sum_{k=0}^{\infty}V_{k}(x,x')\sigma^{k},
\end{equation}
and equating equal powers in $\sigma$ which gives a set of transport equations for each $V_{k+1}(x,x')$ (with $k\ge 0$), namely \cite{Balakumar:2019djw}, 
\begin{subequations}
  \label{eq:Vkeqn}  
\begin{multline}
(k+1)\left[2\sigma^{;\alpha }D_{\alpha }+\Box\sigma+2k\right]V_{k+1}(x,x') \\
+ \left[ D_{\alpha }D^{\alpha }-(\mu^{2}+\xi R)\right] V_{k}(x,x') =0,
\label{eq:Vkeqngen}
\end{multline}
together with the transport equation for $V_{0}$
\begin{multline}
[ 2\sigma^{;\alpha }D_{\alpha }+\Box\sigma-2]V_{0}(x,x') \\
   +[ D_{\alpha }D^{\alpha }-(\mu^{2}+\xi\,R)] U(x,x')=0.
   \label{eq:V0eqn}
\end{multline}
\end{subequations}
The $V_{k}(x,x')$ are now easily solved as covariant Taylor series, subject to boundary conditions obtained by taking the coincidence limits of the respective equations.

We are now in a position to define the renormalized expectation values of the scalar condensate, the current and the stress-energy tensor for the field in the state $|A\rangle$ as follows \cite{Balakumar:2019djw}:
\begin{subequations}
\label{eq:expvalues}
\begin{align}
    \langle |\hat{\Phi}|^{2}\rangle_{A}&=\llbracket  G_{A}(x,x')-K(x,x')\rrbracket\\
    \langle \hat{J}_{\alpha}\rangle_{A}&=-\frac{q}{4\pi}\llbracket \Im\left\{D_{\alpha}\left(G_{A}(x,x')-K(x,x')\right)\right\}\rrbracket\\
    \langle \hat{T}^{\alpha}{}_{\beta}\rangle_{A}&=\llbracket\Re\left\{\hat{\mathsf{t}}^{\alpha}{}_{\beta}\left(G_{A}(x,x')-K(x,x')\right)\right\}\rrbracket+\frac{v_{1}}{4\pi^{2}}\delta^{\alpha}{}_{\beta}
\end{align}
where $\llbracket \cdot\rrbracket$ denotes the coincidence limit,  $\hat{\mathsf{t}}^{\alpha}{}_{\beta}$ is the differential operator
\begin{align}
    \hat{\mathsf{t}}^{\alpha}{}_{\beta}  = & 
    \left(1-2\xi\right)g_{\beta}{}^{\lambda '}D^{\alpha  }D_{\lambda'}^{\dagger}
    +\left(2\xi-\frac{1}{2}\right)\delta^{\alpha }{}_{\beta }g^{\rho\lambda'}D_{\rho}D_{\lambda'}^{\dagger}
    \nonumber\\
& -2\xi D^{\alpha }D_{\beta}
+2\xi\delta^{\alpha }{}_{\beta }D_{\rho}D^{\rho}+\xi R^{\alpha }{}_{\beta }
\nonumber \\ & -\frac{1}{2}(\mu ^{2}+\xi R)\delta^{\alpha  }{}_{\beta },
\end{align}
with $g^{\rho \lambda '}$ the bivector of parallel transport, and
\begin{align}
    v_{1}& =
    \frac{1}{8}\left[ \mu^{2}+(\xi-\frac{1}{6})R\right]^{2}
    -\frac{1}{24}\left(\xi-\frac{1}{5}\right)\Box R
    \nonumber\\
    & \quad 
    -\frac{1}{720}R^{\alpha\beta}R_{\alpha \beta }+\frac{1}{720}R^{\alpha \beta \rho\lambda}R_{\alpha \beta \rho\lambda}
    -\frac{1}{48}q^{2}F^{\alpha \beta }F_{\alpha \beta }
\end{align}
\end{subequations}
is the local term which must be added to ensure the stress-energy tensor is conserved.

For the field in the Hartle-Hawking state $|{\mathrm {H}}\rangle $, one can adopt Euclidean techniques to compute the renormalized expectation values (\ref{eq:expvalues}) \cite{Taylor:2022sly,Breen:2024ggu}. 
Working with the Euclideanized metric has several advantages, for example, the frequency spectrum of the two-point function is discretized on the Euclidean metric and this is computationally more efficient. 
Another advantage is that one does not require an `$i\varepsilon$' prescription [as in (\ref{eq:KHad})] to define the distributions involved since the only singularities in the two-point fucntions and in the Hadamard parametrix are in the coincidence limit; there are no singularities along the null cone since the Euclidean metric is positive definite on the exterior. 
However, the major advantage for working on the Euclidean section is that very fast and efficient mode-sum renormalization prescriptions have been developed for neutral scalar fields on static black hole space-times in the Euclidean context \cite{Taylor:2022sly}, and extended to the charged scalar field in Ref.~\cite{Breen:2024ggu}. 

Here we wish to leverage the efficiency of the Euclidean computation of renormalized quantities in the Hartle-Hawking state in Ref.~\cite{Breen:2024ggu} to develop efficient computational schemes for other states also. 
The key concept involved is that the Hadamard parametrix (\ref{eq:KHad}), which contains the short-distance singularities, is state-independent and therefore the differences of renormalized expectation values computed in different Hadamard states must be finite. 

We therefore consider the difference between two-point functions computed in states $|A\rangle$ and $|B\rangle$,
\begin{align}
   \delta G_{\subAB}(x,x')\equiv G_{A}(x,x')-G_{B}(x,x').
\end{align}
Provided both these states are Hadamard, this difference requires no renormalization, that is, it will be finite at coincidence. 
We can therefore use this fact to express renormalized expectation values in the Boulware and Unruh states in terms of renormalized expectation values in the Hartle-Hawking state plus terms that involve finite differences of two-point functions. 
In particular, let $|{\mathrm {H}}\rangle$ be the Hartle-Hawking state and $|A\rangle$ any other Hadamard state, then
\begin{subequations}
\label{eq:expdiffs}
\begin{align}
\langle |\hat{\Phi}|^{2}\rangle_{A}&=\langle |\hat{\Phi}|^{2}\rangle_{{\mathrm {H}}}+\llbracket\delta G_{\subAH}\rrbracket,\\
 \langle \hat{J}_{\alpha}\rangle_{A}&=\langle \hat{J}_{\alpha}\rangle_{{\mathrm {H}}}-\frac{q}{4\pi}\Im\llbracket (\delta G_\subAH)_{;\alpha} \rrbracket+\frac{q^{2}}{4\pi}A_{\alpha}\llbracket\delta G_\subAH\rrbracket\\
    \langle \hat{T}^{\alpha}{}_{\beta}\rangle_{A}&= \langle \hat{T}^{\alpha}{}_{\beta}\rangle_{{\mathrm {H}}}-\Re\llbracket (\delta G_{\subAH})^{;\alpha}{}_{\beta}\rrbracket
    \nonumber \\ & \quad
    -\left(\xi-\frac{1}{2}\right)\llbracket \delta G_\subAH\rrbracket{}^{;\alpha}{}_{\beta}
    \nonumber \\ & \quad
    +\left(\xi-\frac{1}{4}\right)\delta^{\alpha}{}_{\beta}\Box\llbracket\delta G_{\subAH}\rrbracket+\xi R^{\alpha}{}_{\beta}\llbracket\delta G_{\subAH}\rrbracket\nonumber\\
    & \quad +q^{2}A^{\mu}A_{\nu}\,\llbracket\delta G_{\subAH}\rrbracket-qA^{\mu}\Im\llbracket (\delta G_{\subAH})_{;\nu}\rrbracket\nonumber\\
    &\quad -qA_{\nu}\Im\llbracket (\delta G_{\subAH})^{;\mu}\rrbracket.
\end{align}
\end{subequations}
In arriving at these expressions, we made use of the fact that for any sesquisymmetric biscalar $W(x,x')$, we can prove the following identities:
\begin{subequations}
    \begin{align}
&\Im\llbracket W\rrbracket=0,\\
&\Re\llbracket W_{;\lambda'}\rrbracket=\Re\llbracket W_{;\lambda}\rrbracket=\frac{1}{2}\llbracket W\rrbracket_{;\lambda},\\
   & \Re\llbracket W^{;\nu}{}_{\lambda'}\rrbracket=\frac{1}{2}\llbracket W\rrbracket^{;\nu}{}_{\lambda}-\Re\llbracket W^{;\nu}{}_{\lambda}\rrbracket,\\
   & \Re\{i\,\llbracket W_{;\lambda'}\rrbracket\}=-\Re\{i\,\llbracket W_{;\lambda}\rrbracket\}=\Im\llbracket W_{;\lambda}\rrbracket.
   \\ \nonumber
\end{align}
\end{subequations}

Assuming we can compute the renormalized quantities in the Hartle-Hawking state on the right-hand-side of Eq.~(\ref{eq:expdiffs}), using the methodology of Ref.~\cite{Breen:2024ggu}, then all that remains is to compute the coincidence limit of $\delta G_\subAH$ and its derivatives. For the Boulware and Unruh states, we have the following mode-sum representations of these differences
\begin{widetext}
\begin{subequations}
\begin{align}
    \delta G_{\textrm{\textit{\tiny{B,H}}}}( x,x')&=\frac{1}{16\pi^{2} rr'}\sum_{\ell =0}^{\infty}(2\ell +1)P_{\ell }(\cos\gamma)\int_{-\infty}^{\infty}d\omega\frac{e^{-i\omega(t-t')}}{1-e^{2\pi|\tilde{\omega}|/\kappa}}\left[\frac{\Theta(\omega^{2}-\mu^{2})}{\sqrt{\omega^{2}-\mu^{2}}}X^{\textrm{in}}_{\omega \ell }(r)X^{\textrm{in}*}_{\omega \ell }(r')+\frac{1}{|\tilde{\omega}|}X^{\textrm{up}}_{\omega \ell }(r)X^{\textrm{up}*}_{\omega \ell }(r')\right] ,\\
     \delta G_{\textrm{\textit{\tiny{U,H}}}}( x,x')&=\frac{1}{16\pi^{2} rr'}\sum_{\ell =0}^{\infty}(2\ell +1)P_{\ell }(\cos\gamma)\int_{-\infty}^{\infty}d\omega\frac{e^{-i\omega(t-t')}}{1-e^{2\pi|\tilde{\omega}|/\kappa}}\left[\frac{\Theta(\omega^{2}-\mu^{2})}{\sqrt{\omega^{2}-\mu^{2}}}X^{\textrm{in}}_{\omega \ell }(r)X^{\textrm{in}*}_{\omega \ell }(r')\right].
\end{align}
\end{subequations}
\end{widetext}
These mode-sums are rapidly convergent in $\ell $ and $\omega$ even at coincidence.

\section{Numerical results}
\label{sec:numerics}

We now turn to our numerical results for the renormalized expectation values constructed in the previous section. 
For the Hartle-Hawking state, our numerical methodology is that developed in Ref.~\cite{Breen:2024ggu},
where we presented results for a single choice of the parameters in the theory (namely the black hole charge $Q$; the scalar field mass $\mu $ and charge $q$; and the coupling $\xi $ between the scalar field and the curvature). 
Renormalized expectation values in the Unruh and Boulware states are computed using the state-subtraction method discussed in the previous section.
To implement this numerically, we work on the Lorentzian space-time, and compute the scalar field modes using the method employed in Ref.~\cite{Montagnon:2025vtk}, noting the difficulties inherent for the bound state modes, as discussed in Ref.~\cite{Arrechea:2023fas}.

\subsection{General properties}
\label{sec:gen}

As discussed in Sec.~\ref{sec:Boulware}, there are two quantities which do not require renormalization: $\langle {\hat {J}}^{r} \rangle $ (\ref{eq:Jr}) and $\langle {\hat {T}}^{r}{}_{t}\rangle $ (\ref{eq:Trt}). 
Conservation of the current (\ref{eq:currentcons}) places no constraints on the component $\langle {\hat {J}}^{t}\rangle $, which requires renormalization and therefore must be computed using the methodology outlined in Sec.~\ref{sec:expvalues}.  
For all the states considered here, the remaining components of the current, $\langle {\hat {J}}^{\theta }\rangle $ and $\langle {\hat {J}}^{\varphi }\rangle $, vanish identically \cite{Balakumar:2022yvx,Breen:2024ggu}.

We have already used the $r$-component of the conservation equations (\ref{eq:conservation}) to determine the form of the RSET component $\langle {\hat {T}}^{r}{}_{t}\rangle $ (\ref{eq:Trt}). 
The remaining nontrivial conservation equation arises from setting $\nu = t$ in (\ref{eq:conservation}), and integrating this gives form of the RSET for time-independent states to be \cite{Balakumar:2022yvx}
\begin{equation}
    \langle {\hat {T}}^{\gamma  }{}_{\alpha }\rangle 
    = \left( 
    \begin{array}{cccc}
    {\mathcal {A}}(r) & -f(r)^{-2}\langle {\hat {T}}^{r}{}_{t} \rangle & 0 & 0
    \\
    \langle {\hat {T}}^{r}{}_{t} \rangle & {\mathsf{T}}(r) - {\mathcal {A}}(r) - 2{\mathcal {Q}}(r) & 0 & 0
     \\
     0 & 0 & {\mathcal {Q}}(r) & 0 
     \\
     0 & 0 & 0 & {\mathcal {Q}}(r) 
    \end{array}
    \right) ,
    \label{eq:RSETsol}
\end{equation}
where the trace ${\mathsf{T}}(r)$ is given by \cite{Balakumar:2019djw}
\begin{align}
    {\mathsf{T}}(r) = \langle {\hat {T}}^{\alpha  }{}_{\alpha  } \rangle 
     & = \frac{1}{720\pi ^{2}r^{8}} \left( 13Q^{2} - 24rMQ^{2} + 12r^{2}M^{2} \right)
     \nonumber \\ & \quad 
    - \frac{q^{2}Q^{2}}{96 \pi ^{2}r^{4}}
    -\left[ \mu ^{2}- 3\left(  \xi - \frac{1}{6} \right)\right] \Box \langle {\hat {\Phi }}^{2} \rangle ,
\end{align}
and ${\mathcal {A}}(r)$, ${\mathcal {Q}}(r)$ are state-dependent functions of the radial coordinate $r$ which must be computed numerically. 
The function ${\mathcal {A}}(r)$ can be written as follows:
\begin{align}
    {\mathcal {A}}(r) & = {\mathsf {T}}(r)  - 2{\mathcal {Q}}(r) + \frac{{\mathcal {Z}}}{r^{2}f(r)}
    \nonumber \\ & \qquad 
    + \frac{1}{2r^{2}f(r)} \int _{r_{+}}^{r} \left[ 2{\mathcal {Q}}(r') - {\mathsf {T}}(r') \right]
    r'^{2} f'(r') \, dr'
    \nonumber \\ & \qquad 
    - \frac{1}{2r^{2}f(r)} \int _{r_{+}}^{r} \left[  {\mathcal {Q}}(r') - 
    \frac{2\pi Q}{r'} \langle {\hat {J}}^{t} \rangle \right] r' f(r') \, dr' ,
\end{align}
where ${\mathcal {Z}}$ is an arbitrary constant of integration. 
We note that there is an additional unknown function of $r$ in \eqref{eq:RSETsol} compared with the corresponding analysis for a neutral scalar field (see, for example, \cite{Christensen:1977jc}),
since ${\mathcal {A}}(r)$ depends on the component $\langle {\hat {J}}^{t} \rangle$ which is {\it {a priori}} unconstrained. 

Of particular interest for our later analysis is the behaviour of the renormalized expectation values across the past and future event horizons of the black hole. 
To investigate the regularity of the charged scalar current and RSET at the event horizon, it is helpful to work in Kruskal coordinates $U$, $V$  (\ref{eq:Kruskal}), in terms of which the components of the current are \cite{Balakumar:2022yvx}
\begin{subequations}
\label{eq:expKruskal}
    \begin{align}
    \langle {\hat {J}}^{U} \rangle & =  \frac{\kappa U}{f(r)} \left(  \langle {\hat {J}}_{t} \rangle + \langle {\hat {J}}^{r} \rangle \right) ,
    \\
    \langle {\hat {J}}^{V} \rangle & = -\frac{\kappa V}{f(r)} \left(  \langle {\hat {J}}_{t} \rangle - \langle {\hat {J}}^{r} \rangle \right) ,
\end{align}
while those of the RSET are \cite{Balakumar:2022yvx}
\begin{align}
    \langle {\hat {T}}_{UU} \rangle & =  
   -\frac{f(r)}{4\kappa ^{2}U^{2}} \left( \langle {\hat {T}}^{t}{}_{t} \rangle  -2f(r) \langle {\hat {T}}^{t}{}_{r} \rangle - \langle {\hat {T}}^{r}{}_{r} \rangle  \right) ,
    \\
    \langle {\hat {T}}_{UV} \rangle & =  \frac{f(r)}{4\kappa ^{2}UV} \left( \langle {\hat {T}}^{t}{}_{t} \rangle  + \langle {\hat {T}}^{r}{}_{r} \rangle  \right) ,
    \\
    \langle {\hat {T}}_{VV} \rangle & =  -\frac{f(r)}{4\kappa ^{2}V^{2}} \left( \langle {\hat {T}}^{t}{}_{t} \rangle  +2f(r) \langle {\hat {T}}^{t}{}_{r} \rangle - \langle {\hat {T}}^{r}{}_{r} \rangle  \right)  .
\end{align}
\end{subequations}

\begin{figure*}
    \centering
    \includegraphics[width=0.45\textwidth]{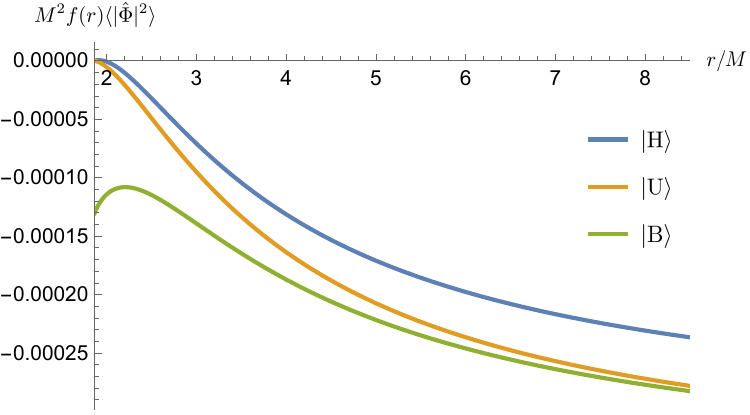}
         \includegraphics[width=0.45\textwidth]{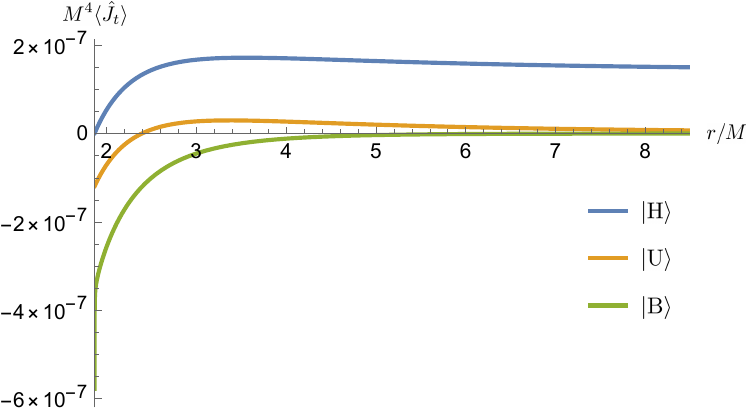}
         \includegraphics[width=0.45\textwidth]{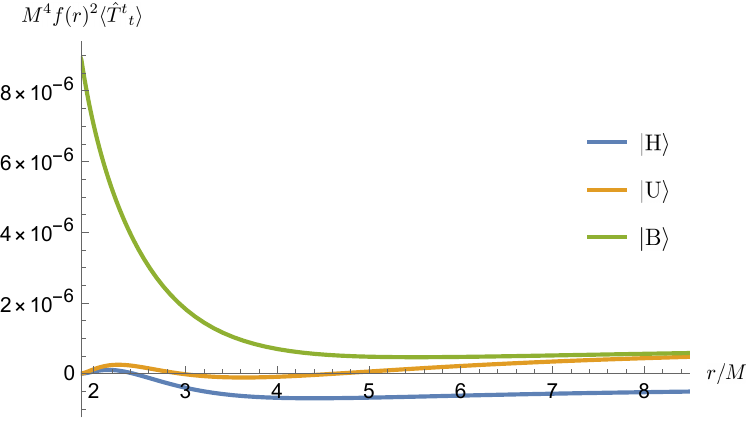}
         \includegraphics[width=0.45\textwidth]{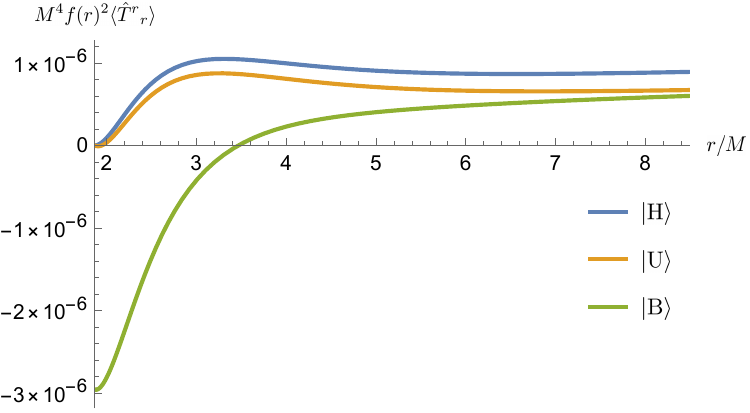}
         \includegraphics[width=0.45\textwidth]{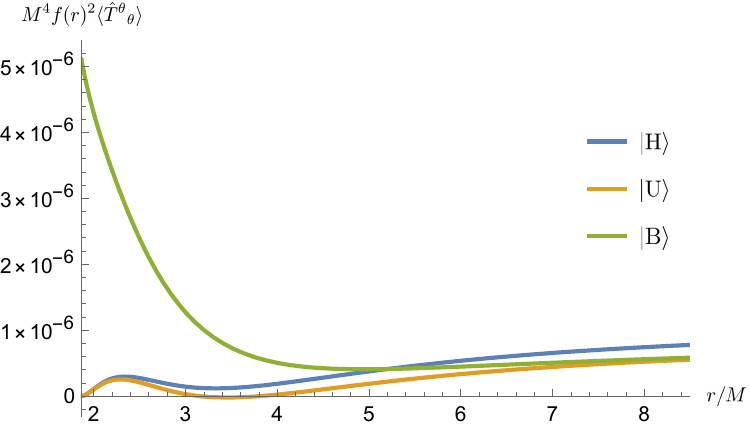}
    \centering
    \caption{Renormalized expectation values for a charged scalar field in the Hartle-Hawking (blue curves), Unruh (orange curves) and Boulware (green curves) states on the RN space-time, with $M=L=1$, $Q=0.5M$, $\xi =0$, $\mu M =1/10$ and $q M =1/4$.}
    \label{fig:threestates}
\end{figure*}

\subsection{Comparing the three states}
\label{sec:ThreeStates}

We begin our discussion of our numerical results by presenting, in Fig.~\ref{fig:threestates}, the renormalized expectation values of the scalar condensate, current and RSET for the Boulware, Unruh and Hartle-Hawking states. 
We have fixed the black hole mass $M=1$, the black hole charge $Q=0.5M$, the scalar field mass $\mu M = 1/10$ and scalar field charge $qM=1/4$.
The charged scalar field is minimally coupled to the curvature, $\xi = 0$.
The remaining parameter in the theory, the renormalization length scale $L$, is also set equal to unity. 
This parameter set is identical to that employed in \cite{Breen:2024ggu}, so the results for the Hartle-Hawking state in Fig.~\ref{fig:threestates} match those in \cite{Breen:2024ggu}. 
We will fix $M=1=L$ throughout this discussion of our numerical results. 

Examining first the scalar condensate (top left in Fig.~\ref{fig:threestates}), we note that, at the horizon, $f(r)\langle | {\hat {\Phi }} |^{2}\rangle _{{\mathrm {B}}}$ is finite and nonzero in the Boulware state, indicating that the scalar condensate diverges like $f(r)^{-1}$ as the horizon is approached. 
This is the same rate of divergence as found for a neutral scalar field on a Schwarzschild black hole \cite{Candelas:1980zt,Levi:2015eea,Levi:2016esr}, and for a massless charged scalar field
on the RN background \cite{Montagnon:2025vtk}.
For both the Unruh and Hartle-Hawking states, the scalar condensate is finite at the horizon, again in agreement with previous results for neutral scalar fields \cite{Candelas:1980zt,Anderson:1990jh}.
For a massless charged scalar field, the scalar condensate is also regular at the horizon in the Unruh state, but the Hartle-Hawking state as considered in this paper for a massive scalar field cannot be constructed in the massless case \cite{Balakumar:2022yvx}. 
Instead, in Ref.~\cite{Montagnon:2025vtk} the CCH (Candelas/Chrzanowski/Howard) state \cite{Candelas:1981zv} is studied, which also has a scalar condensate which is regular at the RN event horizon. 
Far from the black hole, the scalar condensate in all three states is decreasing, and the expectation values in the Boulware and Unruh states tend to the same asymptotic value, while that for the Hartle-Hawking state is different. 
Similar behaviour is observed for a massless charged scalar field \cite{Montagnon:2025vtk}, with the CCH state replacing the Hartle-Hawking state.
It is notable that the scalar condensate presented in Fig.~\ref{fig:threestates} is negative everywhere on and outside the event horizon, while that presented in \cite{Montagnon:2025vtk} for a massless charged scalar field is positive throughout the black hole exterior for the Unruh and CCH states, while, for the Boulware state, it is positive everywhere except for a small neighbourhood of the event horizon.
We shall explore the consequences of varying the black hole and field parameters later in this section.

\begin{figure}
    \centering
    \includegraphics[width=0.45\textwidth]{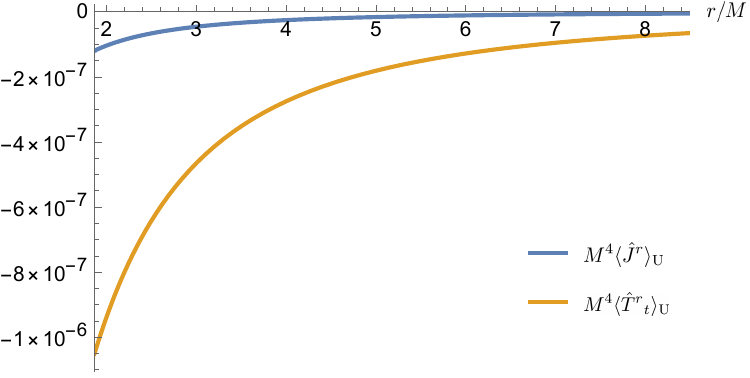}
    \includegraphics[width=0.45\textwidth]{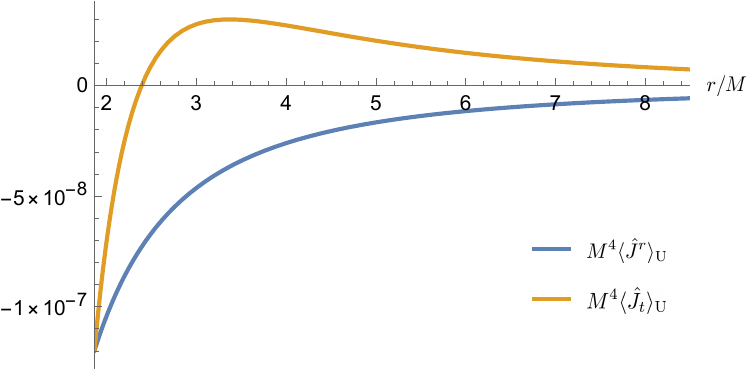}
    \centering
    \caption{Renormalized expectation values of the fluxes (top), and the components of the charged scalar current (bottom) for a charged scalar field in the Unruh state, with $M=L=1$, $Q=0.5M$, $\xi =0$, $\mu M =1/10$ and $q M =1/4$.}
    \label{fig:Unruhfluxes}
\end{figure}

\begin{figure}
\centering
\includegraphics[width=0.45\textwidth]{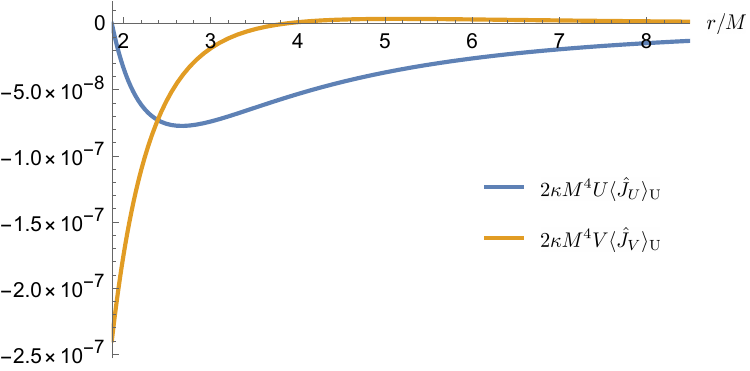}
\centering
\caption{Renormalized expectation values of the components of the charged scalar current in Kruskal coordinates (\ref{eq:expKruskal}) for a charged scalar field in the Unruh state, with $M=L=1$, $Q=0.5M$, $\xi =0$, $\mu M =1/10$ and $q M =1/4$.}
\label{fig:JUnruhKruskal}
\end{figure}

\begin{figure}
    \centering
    \includegraphics[width=0.45\textwidth]{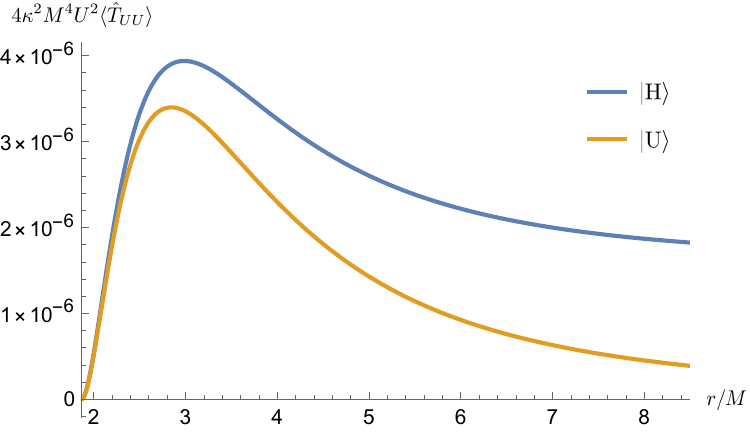}
    \includegraphics[width=0.45\textwidth]{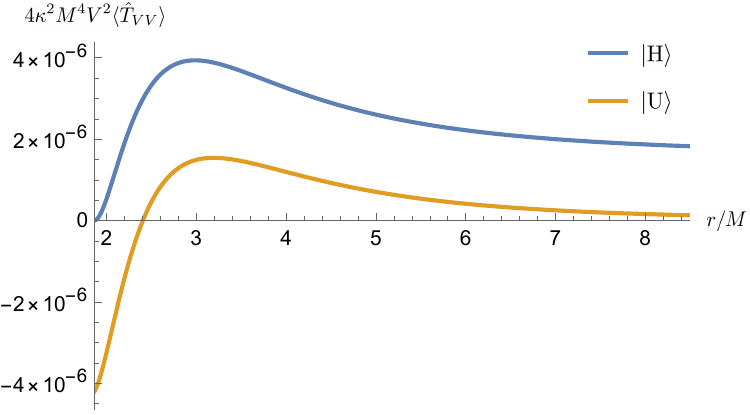}
    \caption{Renormalized expectation values of Kruskal components of the RSET (\ref{eq:expKruskal}) for a charged scalar field in the Hartle-Hawking (blue curves) and Unruh (orange curves) states on the RN space-time, with $M=L=1$, $Q=0.5M$, $\xi =0$, $\mu M =1/10$ and $q M =1/4$.}
    \label{fig:RSETKruskal}
\end{figure}

\begin{figure*}
     \centering
         \includegraphics[width=0.45\textwidth]{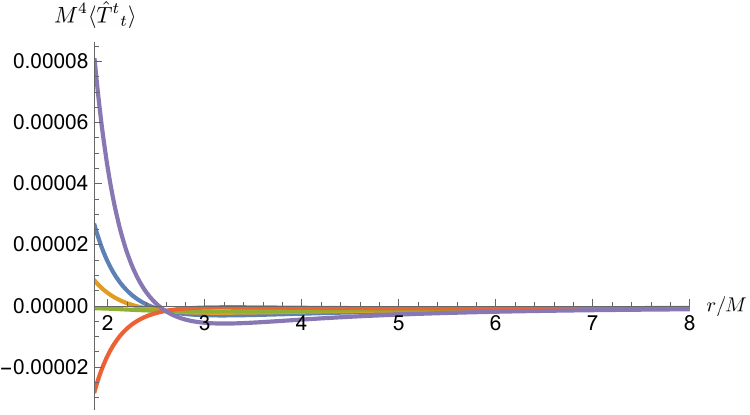}
         \includegraphics[width=0.45\textwidth]{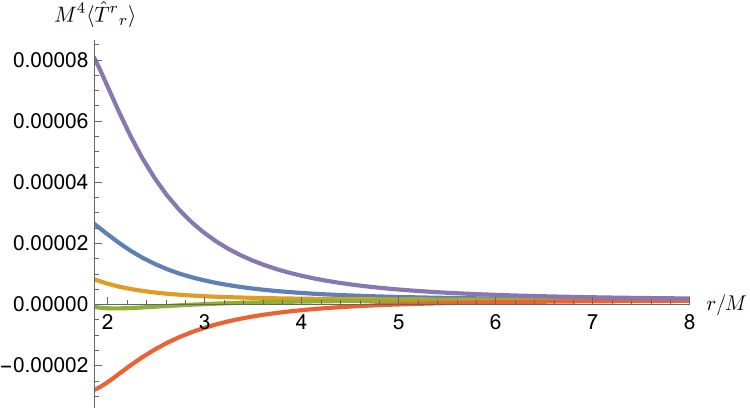}
         \includegraphics[width=0.6\textwidth]{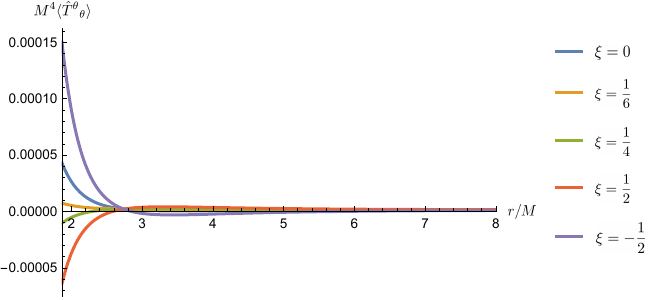}
     \caption{RSET components for a charged scalar field in the Hartle-Hawking state on the RN space-time, with $M=L=1$, $Q=0.5M$, $\mu M =1/10$ and $q M =1/4$, and the coupling constant taking values $\xi=0$ (blue), $1/6$ (orange), $1/4$ (green), $1/2$ (red), $-1/2$ (purple).}
     \label{fig:RSETxi}
\end{figure*}

Next we consider the charged scalar current, presenting $\langle {\hat {J}}_{t}\rangle $ in the top-right plot in Fig.~\ref{fig:threestates}.  
The quantity $\langle {\hat {J}}^{t}\rangle =-[f(r)]^{-1}\langle {\hat {J}}_{t}\rangle $ represents the charge density and vanishes identically for a neutral scalar field. 
For a charged scalar field, the magnitude of $\langle {\hat {J}}_{t}\rangle $ is approximately four orders of magnitude smaller than 
that of the scalar condensate. 
For all three states, we see that $\langle {\hat {J}}_{t}\rangle $ is finite at horizon.
Far from the black hole, the charge density for the Boulware and Unruh states appears to be tending to zero, but not for the Hartle-Hawking state.   
Both these phenomena are again in agreement with the results for a massless charged scalar field \cite{Montagnon:2025vtk} (replacing the Hartle-Hawking state by the CCH state).
At the horizon, the charge density is positive (since $\langle {\hat {J}}_{t}\rangle $ is negative) for both the Boulware and Unruh states (again in agreement with \cite{Montagnon:2025vtk}), but that for the Hartle-Hawking state is zero (whereas, for the CCH state, the charge density is again positive at the horizon \cite{Montagnon:2025vtk}).  
The other notable feature is that $\langle {\hat {J}}_{t}\rangle _{{\mathrm {H}}}$ is positive everywhere outside the horizon in the Hartle-Hawking state, corresponding to a negative charge density.

The radial component of current $\langle {\hat {J}}^{r}\rangle $ (which represents the flux of charge) is zero everywhere for both the Boulware and Hartle-Hawking states. 
This is to be expected for the Hartle-Hawking state \cite{Breen:2024ggu}, which is a thermal equilibrium state.
For the massless charged scalar field considered in Ref.~\cite{Montagnon:2025vtk}, the radial component of the current is not zero in the Boulware state.  
However, for a massless scalar field there is quantum charge superradiance \cite{Gibbons:1975kk,Balakumar:2020gli}, as a consequence of which the (past) Boulware state considered in  Ref.~\cite{Montagnon:2025vtk} is not an equilibrium state.
In contrast, here we consider a scalar field which is sufficiently massive that there there is no charge superradiance \cite{DiMenza:2014vpa}, and in this situation the Boulware state is an equilibrium state (at zero temperature).
For the Unruh state, the radial component of the current is nonvanishing, and is shown in Fig.~\ref{fig:Unruhfluxes}.
We see that $\langle {\hat {J}}^{r}\rangle _{{\mathrm {U}}}$ is negative everywhere outside the event horizon, corresponding to the black hole losing charge, as expected. 
This is also the case for all three states (Boulware, Unruh and CCH) when the scalar field is massless \cite{Montagnon:2025vtk}. 

The components of the charge current in Kruskal coordinates are shown in Fig.~\ref{fig:JUnruhKruskal} for the Unruh state. 
We see that  $U\langle {\hat {J}}_{U} \rangle $ vanishes as $r\rightarrow r_{+}$, while $V\langle {\hat {J}}_{V} \rangle $ is finite  and nonzero in this limit. 
Therefore the current is regular on the future event horizon (where $U=0$) but diverges on the past event horizon (where $V=0$), in accordance with expectations for the Unruh state.
Since $\langle {\hat {J}}^{r}\rangle $ vanishes identically for both the Boulware and Hartle-Hawking states, the current will only be regular at both the past and future event horizons if $\langle {\hat {J}}_{t} \rangle $ vanishes there. 
From the top-right plot in Fig.~\ref{fig:threestates} we deduce that the current is regular on both horizons in the Hartle-Hawking state, but divergent on both horizons in the Boulware state, again as expected.

We now turn to the RSET, whose nonzero components are shown in the remaining three plots in Fig.~\ref{fig:threestates}, in each case multiplied by a factor of $f(r)^{2}$.  
All three quantities, $f(r)^{2}\langle {\hat {T}}^{t}{}_{t}\rangle$, $f(r)^{2}\langle {\hat {T}}^{r}{}_{r}\rangle$ and $f(r)^{2}\langle {\hat {T}}^{\theta }{}_{\theta } \rangle$, are finite everywhere on and outside the event horizon.
They have similar magnitudes, each roughly an order of magnitude larger than the components of the current, and two orders of magnitude smaller than the scalar condensate. 
In the Unruh state, there is an additional nonzero component of the RSET, namely $\langle {\hat {T}}^{r}{}_{t}\rangle $, which is shown in Fig.~\ref{fig:Unruhfluxes}. 
This is negative everywhere on and outside the horizon, corresponding to an outgoing flux of energy from the black hole, which is then losing mass, again in line with our expectations.

Far from the black hole, all three diagonal components of the RSET are nonzero in all three quantum states, with the curves for the Boulware and Unruh states converging as $r\rightarrow \infty $, while that for the Hartle-Hawking state has a different asymptotic value.  
For the Hartle-Hawking state, the fact that the RSET components approach finite, nonzero constants far from the black hole is the same as happens for a minimally-coupled neutral scalar field \cite{Levi:2016paz,Arrechea:2023fas}.
This is in accordance with our intuition that the Hartle-Hawking state is a thermal equilibrium state. 
For a minimally-coupled neutral scalar field, the RSET in the Unruh and Boulware states tends to zero far from the black hole \cite{Anderson:1994hg,Levi:2016esr,Arrechea:2023fas}, with that in the Boulware state decreasing more rapidly [$\sim {\mathcal {O}}(r^{-4})$] than that in the Unruh state [$\sim {\mathcal {O}}(r^{-2})$].
The difference in behaviour seen for a charged scalar field in Fig.~\ref{fig:threestates} in the Unruh and Boulware states is due to the fact that, even in the absence of superradiance, the electrostatic field surrounding the black hole will spontaneously produce scalar field quanta via Schwinger pair-production (see, for example, \cite{Gibbons:1975kk,Gabriel:2000mg,Khriplovich:1999gm,Khriplovich:1999qa,Khriplovich:2002qn,Chen:2012zn,Johnson:2019kda}).

Near the event horizon, the components of the RSET in the Boulware state diverge like $f(r)^{-2}$ as $r\rightarrow r_{+}$, the same behaviour as found for a neutral scalar field \cite{Levi:2016esr,Arrechea:2023fas,Arrechea:2024cnv}. 
To examine the regularity of the RSET near the horizon in the Unruh and Hartle-Hawking states, we show the Kruskal components of the RSET (\ref{eq:expKruskal}) in Fig.~\ref{fig:RSETKruskal}. 
We see that while $U^{2}\langle {\hat {T}}_{UU}\rangle $ vanishes on the horizon in both these states, in accordance with our expectation that both states are regular on the future event horizon ${\mathcal {H}}^{+}$, the quantity $V^{2}\langle {\hat {T}}_{VV} \rangle $ is nonzero in the limit $r\rightarrow r_{+}$ in the Unruh state.
We deduce that the Unruh state is not regular on the past event horizon ${\mathcal {H}}^{-}$, again as anticipated.
The Hartle-Hawking state, on the other hand, is expected to be regular on both the past and future horizons, as $V^{2}\langle {\hat {T}}_{VV} \rangle $ tends to zero as $r\rightarrow r_{+}$.

The component $-\langle {\hat {T}}^{t}{}_{t}\rangle $ of the RSET is of particular interest, since it corresponds to the energy density in the quantum field. 
Far from the black hole the energy density shown in Fig.~\ref{fig:threestates} is negative in the Unruh and Boulware states, but positive for the Hartle-Hawking state.
The analysis in Ref.~\cite{Arrechea:2023fas} of the RSET for a neutral scalar field on RN space-time shows that whether or not the energy density is positive depends not only on the quantum state under consideration, but also the parameters of the background black hole space-time and the quantum scalar field.
We therefore next explore the effect of changing the parameters $Q$, $\mu $, $q$ and $\xi $ on the renormalized expectation values.  Since the Boulware state is divergent at the event horizon, from here on we study only the Hartle-Hawking and Unruh states. 
We pay particular attention to the Hartle-Hawking state, and discuss the features of the expectation values in the Unruh state where these are qualitatively different from those in the Hartle-Hawking state. 

\begin{figure*}
\centering
         \includegraphics[width=0.45\textwidth]{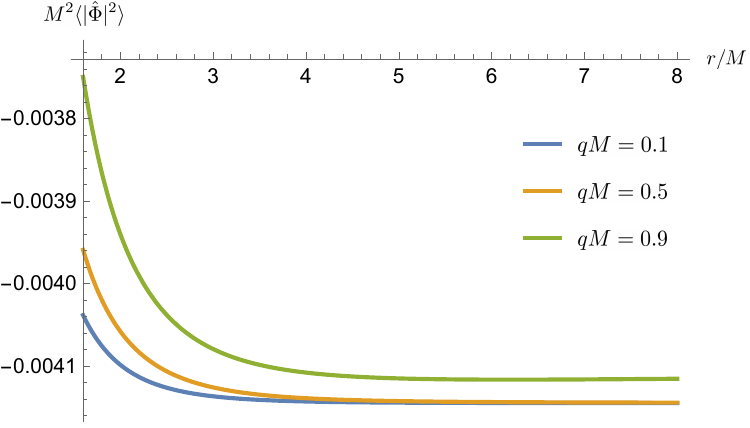}
         \includegraphics[width=0.45\textwidth]{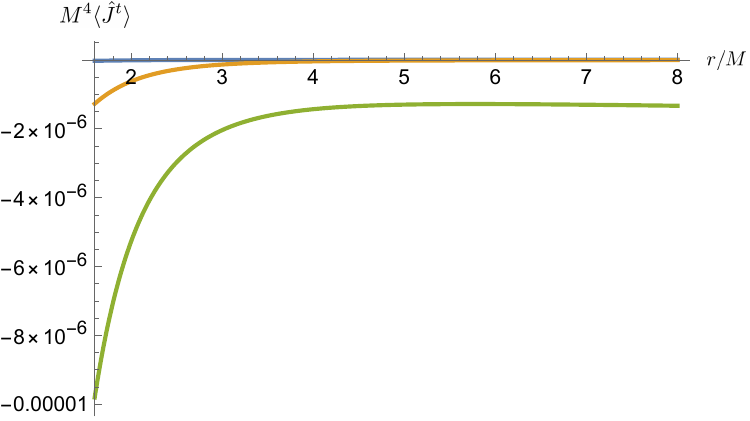}
         \includegraphics[width=0.45\textwidth]{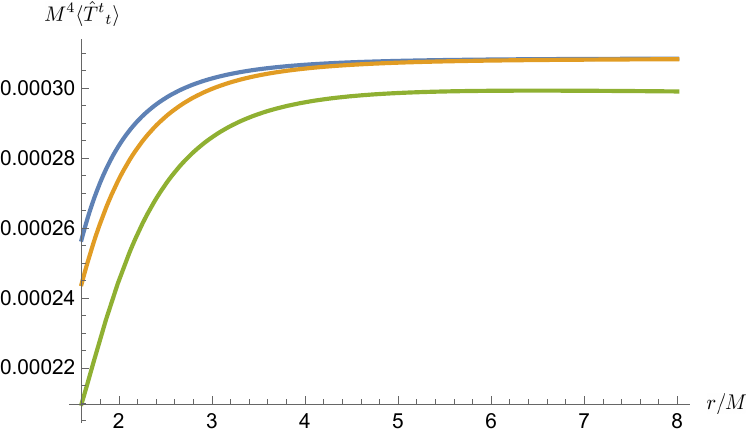}
         \includegraphics[width=0.45\textwidth]{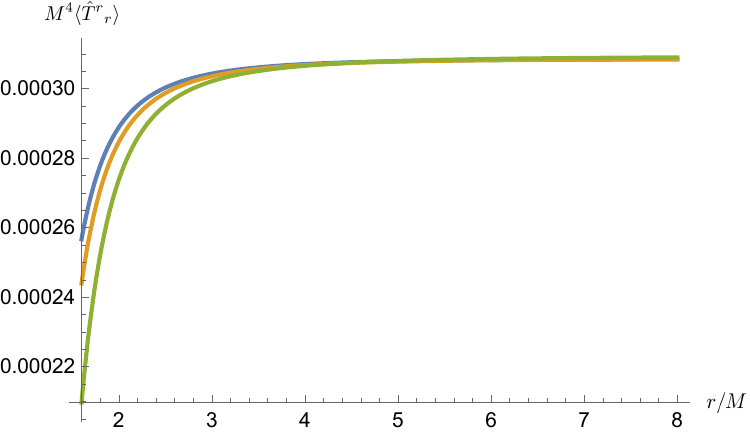}
         \includegraphics[width=0.45\textwidth]{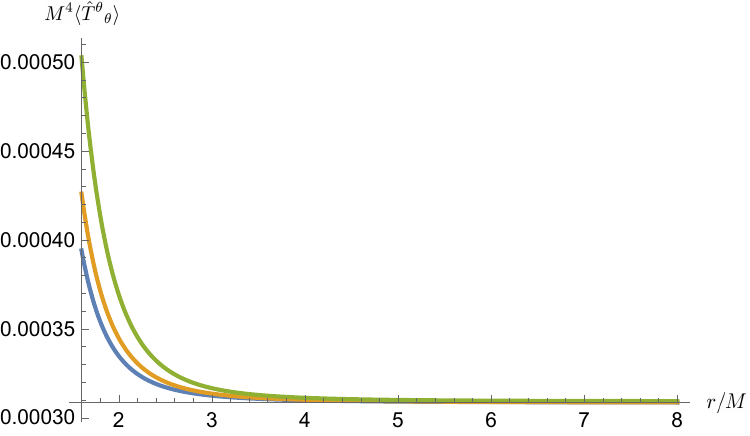}
     \caption{Renormalized expectation values for a charged scalar field on the RN space-time in the Hartle-Hawking state, with $M=L=1$, $Q/M=0.8$, $\xi=0$, $\mu M=1/2$ and the scalar field charge taking values $qM = 0.1$ (blue), $0.5$ (orange), $0.9$ (green).}
     \label{fig:ObsDifferentsq}
\end{figure*}

\begin{figure}
    \centering
    \includegraphics[width=0.45\textwidth]{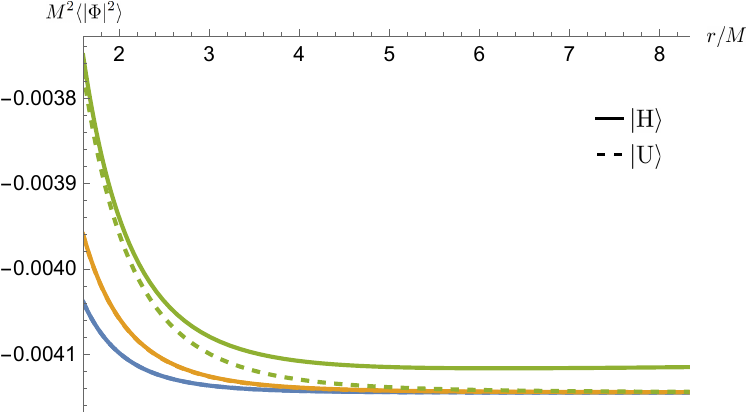}
    \includegraphics[width=0.45\textwidth]{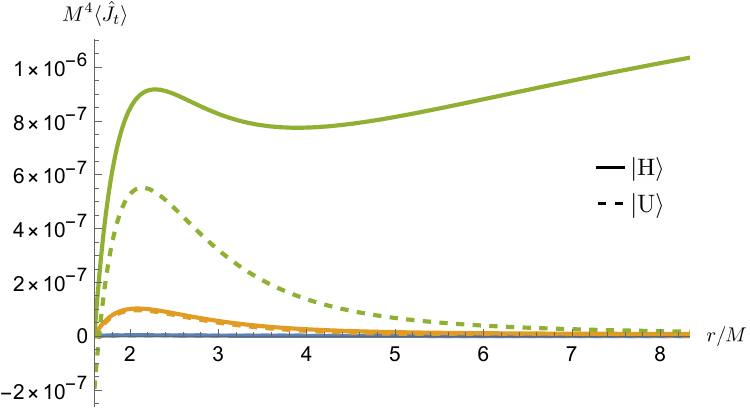}
    \includegraphics[width=0.45\textwidth]{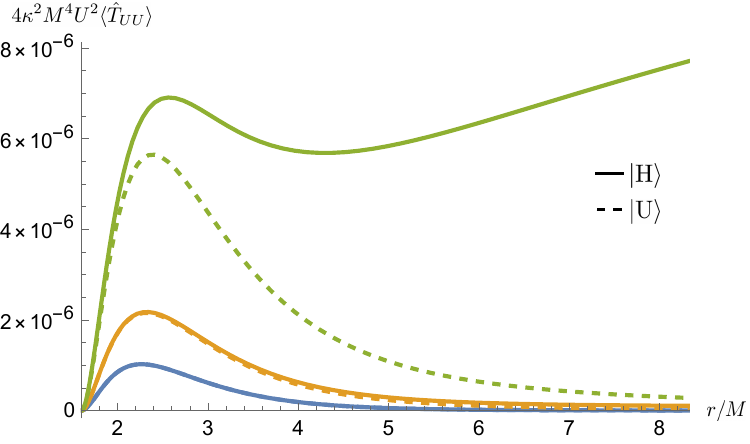}
    \caption{Renormalized expectation values for a charged scalar field on the RN space-time in the Hartle-Hawking (solid lines) and Unruh (dashed lines) states, with $M=L=1$, $Q/M=0.8$, $\xi=0$, $\mu M=1/2$ and the scalar field charge taking values $qM = 0.1$ (blue), $0.5$ (orange), $0.9$ (green). Curves for the components of the renormalized expectation values not shown are almost indistinguishable between the two states.}
    \label{fig:ObsDifferentsqHU}
\end{figure}

\subsection{Varying the coupling}
\label{sec:coupling}

First we consider the effect of varying the parameter $\xi $ which governs the coupling of the scalar field to the scalar curvature (\ref{eq:scalar}).  
Since the Ricci scalar for a background Reissner-Nordstr\"om black hole is zero, varying $\xi $ does not affect the scalar field modes or the renormalized scalar condensate or charge current expectation values. 
Even so, $\xi $ appears explicitly in the classical stress-energy tensor (\ref{eq:SET}) and thus does affect the RSET. 
In Fig.~\ref{fig:RSETxi} we show the nonzero components of the RSET in the Hartle-Hawking state, fixing the black hole charge $Q=0.5M$, and the scalar field mass $\mu = 1/(10M)$ and charge $q=1/(4M)$ to be the same as in Fig.~\ref{fig:threestates}, but varying the coupling constant $\xi $.

The effect of varying $\xi $ on the RSET components is very similar to the situation for a neutral scalar field \cite{Taylor:2022sly,Arrechea:2023fas}.  
In particular, varying $\xi $ has the most significant effect in the region close to the event horizon, and changing $\xi $ does not affect the values of the RSET components far from the black holes. 
As $\xi $ increases, the values of all three RSET components at the event horizon decrease.
When $\xi $ is either negative or positive but sufficiently small, the RSET components are decreasing as the radial coordinate $r$ increases above $r_{+}$, but for sufficiently large and positive $\xi $, they are increasing on moving away from the horizon.  
The other key feature is that the values of the RSET components on the horizon are positive for negative or sufficiently small and positive $\xi $, but become negative for sufficiently large and positive $\xi $. 
Since all these qualitative features match those in the neutral case ~\cite{Arrechea:2023fas}, we do not consider the effect of varying $\xi $ on the RSET components in the Unruh state, since these will likely also mimic those for a neutral scalar field, as studied in Ref.~\cite{Arrechea:2023fas}.  
For the rest of our discussion of the renormalized expectation values, we return to the minimally coupled case with $\xi  =0$.

\begin{figure}
    \centering
    \includegraphics[width=0.45\textwidth]{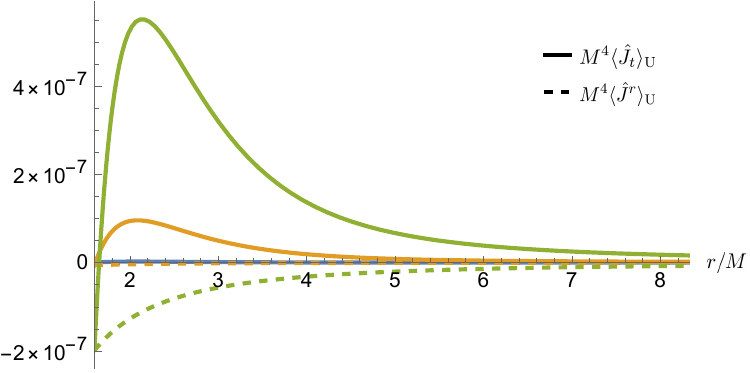}
    \includegraphics[width=0.45\textwidth]{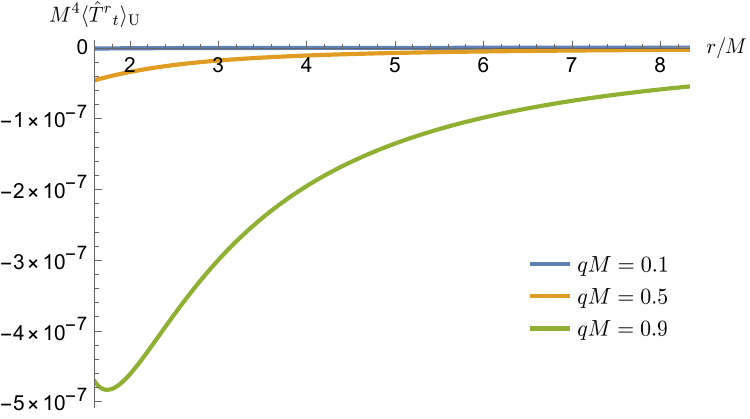}
    \caption{Renormalized expectation values of the charged scalar current (top) and energy flux (bottom) for a charged scalar field on the RN space-time in the Unruh state, with $M=L=1$, $Q/M=0.8$, $\xi=0$, $\mu M=1/2$ and the scalar field charge taking values $qM = 0.1$ (blue), $0.5$ (orange), $0.9$ (green).}
    \label{fig:FluxesDifferentsq}
\end{figure}

\begin{figure*}
\centering
         \includegraphics[width=0.45\textwidth]{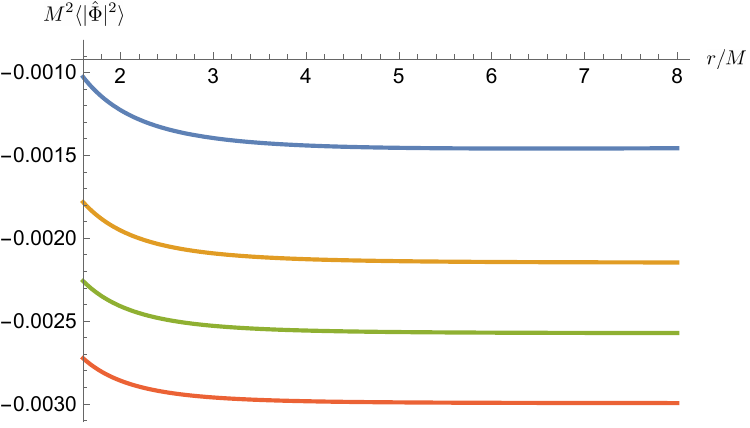}
         \includegraphics[width=0.45\textwidth]{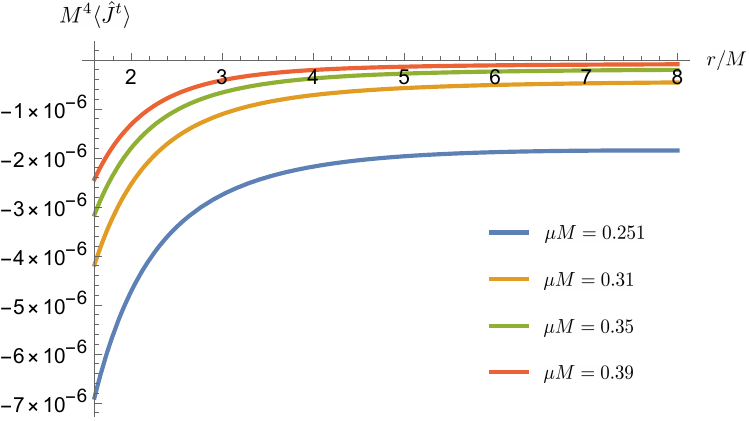}
         \includegraphics[width=0.45\textwidth]{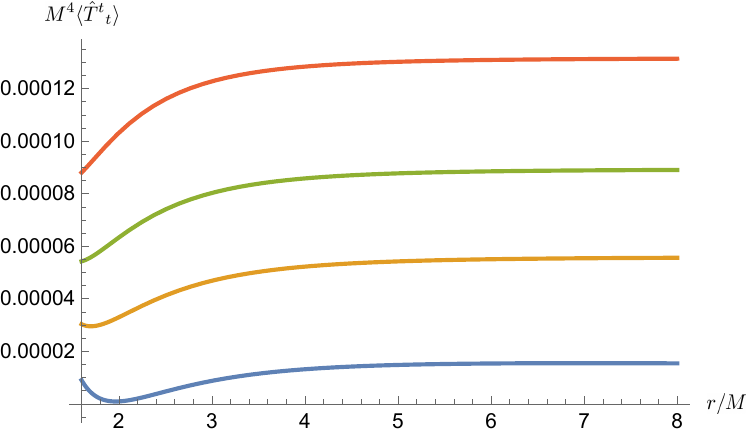}
         \includegraphics[width=0.45\textwidth]{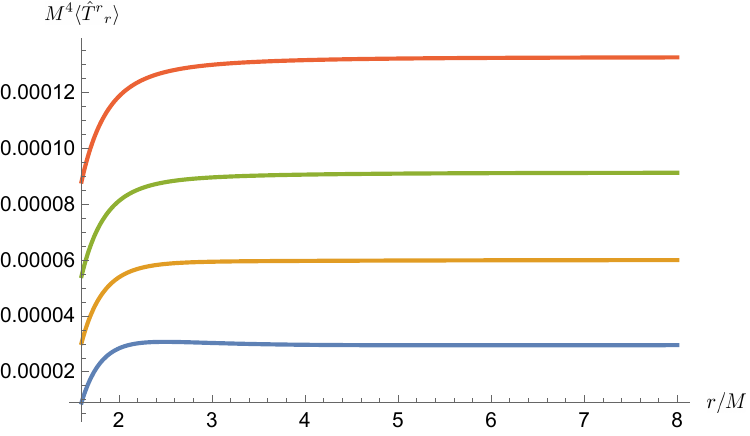}
         \includegraphics[width=0.45\textwidth]{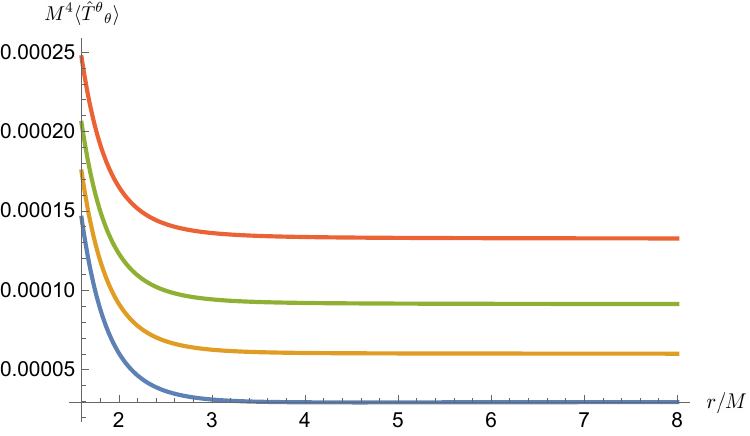}
     \caption{Renormalized expectation values of all nonzero observables for a charged scalar field on the RN space-time in the Hartle-Hawking state, with $M=L=1$, $Q/M=0.8$, $\xi=0$, $q M=1/2$ and the scalar mass taking values $\mu M = 0.251$ (blue), $0.31$ (orange), $0.35$ (green), $0.39$ (red).}
     \label{fig:ObsDifferentMu}
\end{figure*}

\subsection{Varying the scalar field charge}
\label{sec:ScalarCharge}

Next we keep the black hole charge $Q=0.8M$ fixed (note that this is larger than the value considered in Figs.~\ref{fig:threestates}--\ref{fig:RSETxi}) and examine the effect of varying the scalar field parameter $qM$, while considering only a minimally coupled scalar field with $\xi =0$ and fixing the scalar field mass to be $\mu = 1/(2M)$ (this is a larger mass than in our earlier discussions). 

First, in Fig.~\ref{fig:ObsDifferentsq}, we present all the nonzero components of the scalar condensate, charge current and RSET in the Hartle-Hawking state with three different values of the scalar charge $qM=0.1$, $0.5$ and $0.9$. 
In all cases, changing $q$ does not change the general features of the curves for the renormalized expectation values (in particular, whether the quantities are monotonically increasing or decreasing).

For the scalar condensate, increasing the scalar field charge increases the scalar condensate for all values of $r$ (and thus decreasing its magnitude). 
For a massless charged scalar field in the CCH state \cite{Montagnon:2025vtk} (the state closest to the Hartle-Hawking state considered in our present analysis), it is also found that increasing the scalar field charge increases the scalar condensate (there called the ``vacuum polarization''), but there the scalar condensate for the CCH state is positive for the values of the black hole and scalar field charge considered.

The most significant change in magnitude occurs for the time component $\langle {\hat {J}}^{t}\rangle _{{\mathrm {H}}}$ of the charge current, whose value on the event horizon increases by roughly an order of magnitude on increasing $qM$ from $0.5$ to $0.9$.  
Since this quantity vanishes identically for a neutral scalar field, this behaviour is not unexpected.
A similar increase in magnitude of the charge density is observed for a massless charged scalar field \cite{Montagnon:2025vtk}.  
Note that, since $\langle {\hat {J}}^{t}\rangle = -f(r)^{-1}\langle {\hat {J}}_{t}\rangle$, the charge density for the Hartle-Hawking state in Fig.~\ref{fig:ObsDifferentsq} has the same (negative) sign as that in Fig.~\ref{fig:threestates}.

In contrast, the RSET components change rather less significantly on increasing the scalar field charge, particularly the $\langle {\hat {T}}^{r}{}_{r}\rangle _{{\mathrm {H}}}$ and $\langle {\hat {T}}^{\theta }{}_{\theta }\rangle _{{\mathrm {H}}}$ components. 
At the horizon, the value of the energy density $-\langle {\hat {T}}^{t}{}_{t}\rangle _{{\mathrm {H}}}$ is negative for all three values of $qM$ studied, and its magnitude decreases as $qM$ is increased.  

For the parameter values studied in Fig.~\ref{fig:ObsDifferentsq}, most of the renormalized expectation values for the Unruh state are very similar to those for the Hartle-Hawking state. 
In Fig.~\ref{fig:ObsDifferentsqHU} we show only those components for which there is a difference in the expectation values between these two states which can be seen on the plots. 
In Fig.~\ref{fig:ObsDifferentsqHU}, it is only for the very large value of the scalar field charge, $qM=0.9$, for which these differences are significant.
For this value of $qM$, the most striking difference is the behaviour of the components far from the black hole.
This is to be expected, since the Hartle-Hawking state corresponds to a nonempty, thermal equilibrium state, whereas, at infinity, the Unruh state contains only the outgoing flux of Hawking radiation.
What is perhaps more surprising is that the renormalized expectation values in these two states are so similar for the other values of $qM$.
In Figs.~\ref{fig:ObsDifferentsq} and \ref{fig:ObsDifferentsqHU}, we have chosen a large value for the scalar field mass, namely $\mu M = 1/2$, to ensure that the inequality (\ref{eq:SRbound}) is satisfied for all values of $qM$ shown. 
However, when the scalar field mass is large, the renormalized expectation values are dominated by the background curvature and not the quantum state, and as a consequence the expectation values in the Unruh and Hartle-Hawking states differ only by very small amounts.
There is an appreciable difference in the expectation values in these two states only when the scalar field charge is very large, in which case the black hole will still discharge even though the effects of Hawking radiation are very small.

Varying the scalar field charge $qM$ has a particularly significant effect on the charge flux $\langle {\hat {J}}^{r}\rangle $ and the energy flux $\langle {\hat {T}}^{r}{}_{t}\rangle $, both of which vanish identically for the Hartle-Hawking state and are shown for the Unruh state in Fig.~\ref{fig:FluxesDifferentsq} (as well as the charge current component $\langle {\hat {J}}_{t}\rangle _{{\mathrm {U}}}$ for comparison). 
Both fluxes ${\mathcal {K}}_{\mathrm {U}}$ and ${\mathcal {L}}_{\mathrm {U}}$ (\ref{eq:Ufluxes}) increase rapidly with increasing $qM$ (as is the case also for a massless charged scalar field \cite{Montagnon:2025vtk}), demonstrating that, even though the scalar field mass $\mu M$ is large and thus Hawking radiation is very small, the black hole will emit increasing amounts of mass and charge as the scalar field charge increases. 
Thus, for a charged scalar field, discharge effects can remain appreciable even when Hawking radiation is not. 

\subsection{Varying the scalar field mass}
\label{sec:ScalarMass}

Fixing the scalar field charge $qM=1/2$ (as well as $Q/M=0.8$ and $\xi = 0$) and varying the scalar field mass $\mu M$, our results for the renormalized expectation values in the Hartle-Hawking state are shown in Fig.~\ref{fig:ObsDifferentMu}.  
In order to ensure that the bound (\ref{eq:SRbound}) is satisfied for all values of $\mu M $ studied, we have taken a smaller value of the scalar field charge, namely $qM=1/2$, for which (\ref{eq:SRbound}) becomes $\mu M>0.25$ (we have set $M=1$ and $Q/M=0.8$, giving $r_{+}=1.6$).
For values of $\mu M$ below this bound the charged scalar field will exhibit superradiance.

From Fig.~\ref{fig:ObsDifferentMu}, we see that varying $\mu M$ does not significantly change the qualitative shape of the curves for the renormalized expectation values, but generally shifts these curves either up or down.
For a neutral scalar field on an RN black hole \cite{Arrechea:2023fas}, it was also found that changing the scalar field mass did not have a significant effect on the RSET, although only a small nonzero scalar field mass was studied.
On a Schwarzschild black hole, considering a neutral scalar field with larger mass \cite{Taylor:2022sly} has a more significant effect on the magnitude of the RSET component $\langle {\hat {T}}^{r}{}_{r}\rangle _{{\mathrm {H}}}$,  but again does not substantially alter its qualitative profile as a function of $r$. 
The magnitudes of the components shown in Fig.~\ref{fig:ObsDifferentMu} typically decrease as $\mu M$ decreases (in accordance with the results for $\langle {\hat {T}}^{r}{}_{r}\rangle _{{\mathrm {H}}}$ in Ref.~\cite{Taylor:2022sly}), with the notable exception of the charge density $\langle {\hat {J}}^{t}\rangle _{{\mathrm {H}}}$, whose magnitude increases as $\mu M$ approaches the superradiant limit. 
We also notice that the dependence of $\langle {\hat {T}}^{t}{}_{t}\rangle _{{\mathrm {H}}}$ on the radius $r$ changes as $\mu M$ decreases: for larger values of the scalar field mass this quantity is monotonically increasing as $r$ increases, while for $\mu M$ close to the superradiant bound a minimum develops close to the event horizon.

\begin{figure}
    \centering
    \includegraphics[width=0.45\textwidth]{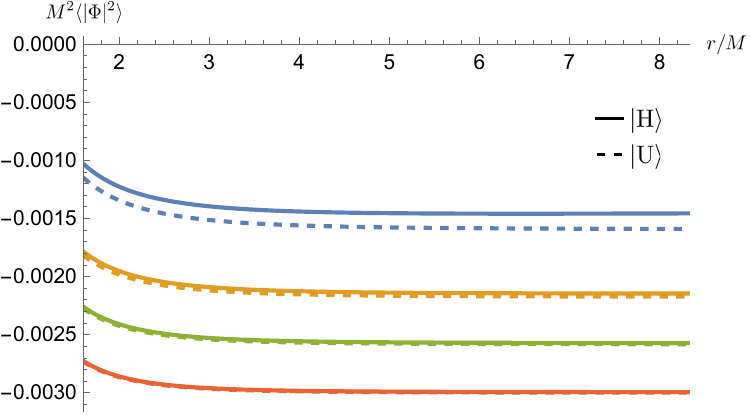}
    \includegraphics[width=0.45\textwidth]{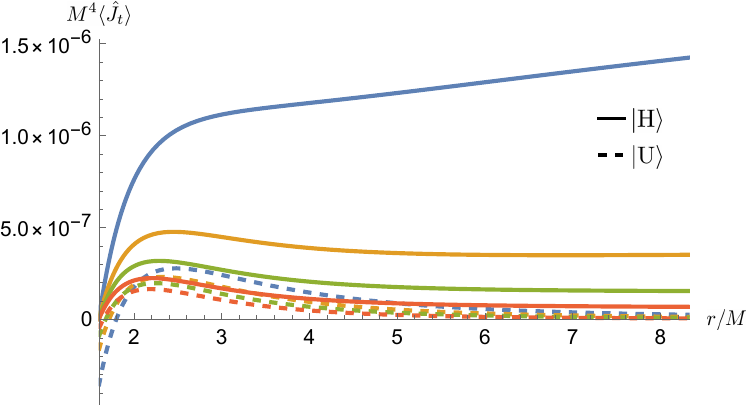}
    \includegraphics[width=0.45\textwidth]{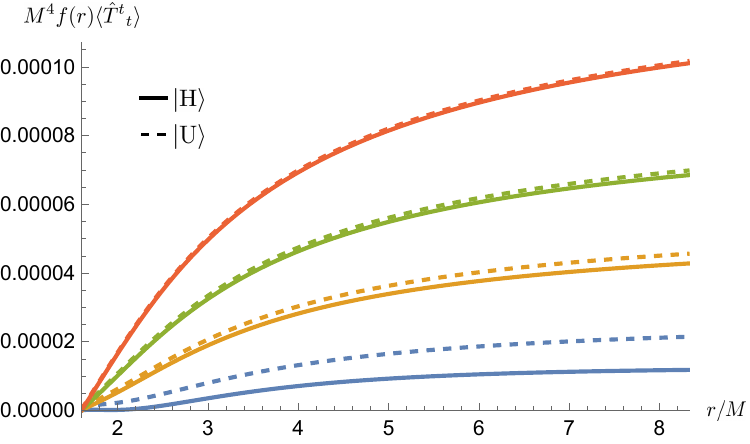}
    \caption{Renormalized expectation values for a charged scalar field on the RN space-time in the Hartle-Hawking (solid lines) and Unruh (dashed lines) states, with $M=L=1$, $Q/M=0.8$, $\xi=0$, $q M=1/2$ and the scalar mass taking values $\mu M = 0.251$ (blue), $0.31$ (orange), $0.35$ (green), $0.39$ (red). Curves for the components of the renormalized expectation values not shown are almost indistinguishable between the two states.}
    \label{fig:ObsDifferentMuHU}
\end{figure}

While the values of the scalar field mass $\mu M$ considered in Fig.~\ref{fig:ObsDifferentMu} are smaller than the value $\mu M =1/2$ in Figs.~\ref{fig:ObsDifferentsq}--\ref{fig:FluxesDifferentsq}, we find it is still the case that many of the components of the renormalized expectation values in the Unruh state, for the parameters used in Fig.~\ref{fig:ObsDifferentMu}, are very similar to those in the Hartle-Hawking state. 
In Fig.~\ref{fig:ObsDifferentMuHU} we focus on those quantities for which the differences between the Unruh and Hartle-Hawking states can be seen in the plots.  
For the scalar condensate $\langle | {\hat {\Phi }}|^{2}\rangle $ and the RSET component $\langle {\hat {T}}^{t}{}_{t}\rangle $, the difference between the two states increases as the scalar field mass decreases, and the effect of Hawking radiation becomes more significant, although the qualitative shapes of the curves do not vary much. 
The magnitudes of both these quantities are slightly larger in the Unruh state compared to the Hartle-Hawking state.

The centre plot in Fig.~\ref{fig:ObsDifferentMuHU} shows the component $\langle {\hat {J}}_{t}\rangle $ of the charge current, whose values in the Unruh and Hartle-Hawking states are rather different, particularly far from the black hole.
At the event horizon,  $\langle {\hat {J}}_{t}\rangle _{\mathrm {H}}$ vanishes in the Hartle-Hawking state (as expected since this state is anticipated to be regular at the horizon), while for the Unruh state, as observed previously, this quantity is finite but nonzero. 
Far from the event horizon, we see that $\langle {\hat {J}}_{t}\rangle _{{\mathrm {U}}}$ tends to zero in the Unruh state, but for the Hartle-Hawking state it either approaches a nonzero constant (as seen previously, in Figs.~\ref{fig:threestates} and \ref{fig:ObsDifferentsqHU}), or, as the superradiant limit approaches, appears to increase with increasing $r$ for all values of the radial coordinate considered.
In Fig.~\ref{fig:ObsDifferentsqHU}, we found for $\mu M=1/2$ and $qM=0.9$ (which is again close to the superradiant bound) that the charge current component $\langle {\hat {J}}_{t}\rangle _{\mathrm {H}}$ for the Hartle-Hawking state was also increasing with increasing radial coordinate sufficiently far from the black hole.
While the Unruh state is well-defined for scalar field masses both below and above the superradiant bound (\ref{eq:SRbound}), the Hartle-Hawking state is defined only when (\ref{eq:SRbound}) is satisfied. 
Hence, in Figs.~\ref{fig:ObsDifferentsqHU} and \ref{fig:ObsDifferentMuHU} we may be seeing indications that the Hartle-Hawking state becomes ill-defined in the superradiant limit. 

\begin{figure}
    \centering
    \includegraphics[width=0.45\textwidth]{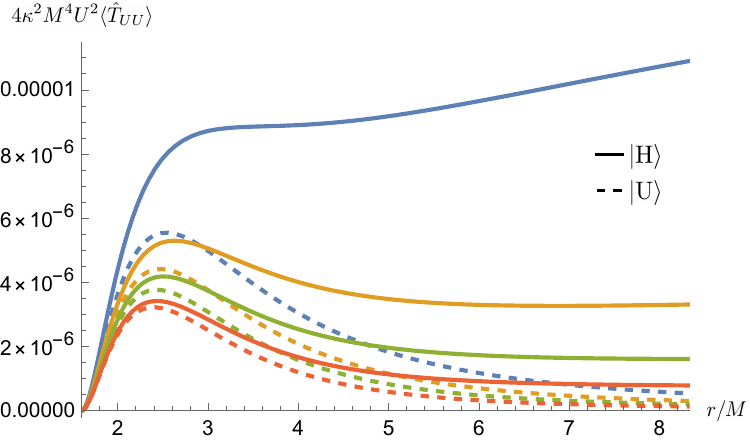}
    \includegraphics[width=0.45\textwidth]{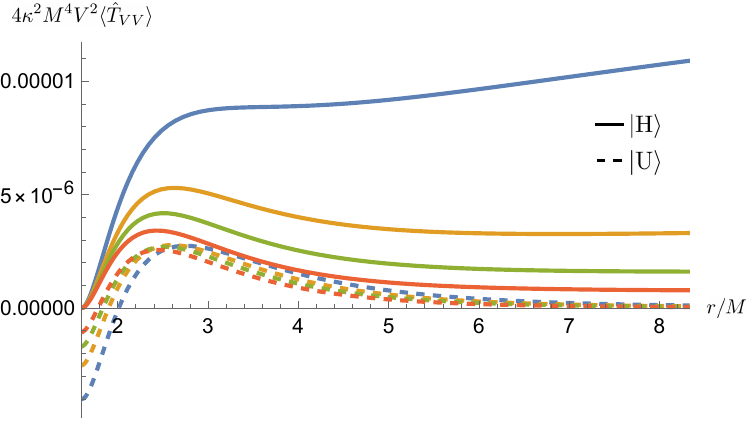}
    \caption{Renormalized expectation values of the RSET in Kruskal coordinates (\ref{eq:expKruskal}) for a charged scalar field on the RN space-time in the Hartle-Hawking and Unruh states, with $M=L=1$, $Q/M=0.8$, $\xi=0$, $q M=1/2$ and the scalar mass taking values $\mu M = 0.251$ (blue), $0.31$ (orange), $0.35$ (green), $0.39$ (red). }
    \label{fig:ObsDifferentMuKruskal}
\end{figure}

We also find appreciable differences between the Hartle-Hawking and Unruh states for the expectation values of the RSET in Kruskal coordinates (\ref{eq:expKruskal}), as shown in Fig.~\ref{fig:ObsDifferentMuKruskal}.
Far from the black hole, both Kruskal components of the RSET tend to zero in the Unruh state, but, for the Hartle-Hawking state, the quantities plotted either approach a finite nonzero limit, or, for sufficiently small mass close to the superradiant bound, are monotonically increasing functions of the radial coordinate $r$.
At the horizon, while $U^{2}\langle {\hat {T}}_{UU} \rangle $ tends to zero for both quantum states, we find that $V^{2}\langle {\hat {T}}_{VV}\rangle $ tends to zero in the Hartle-Hawking state, but remains finite and nonzero in the Unruh state.
This is in accordance with our expectations that the Unruh state is regular on the future but not the past event horizon, while the Hartle-Hawking state is regular on both the future and the past horizon.

\begin{figure}
    \centering
    \includegraphics[width=0.45\textwidth]{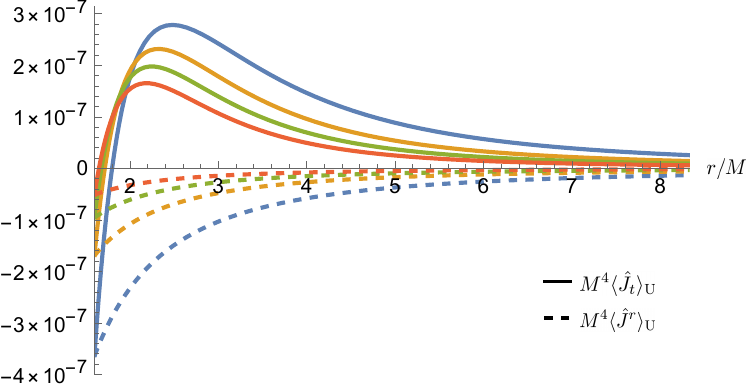}
    \includegraphics[width=0.45\textwidth]{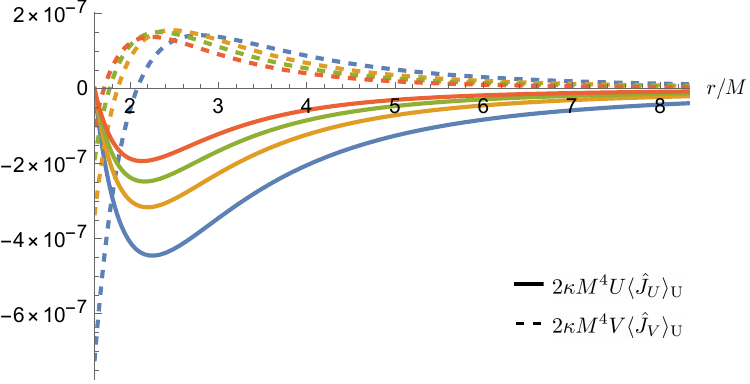}
    \includegraphics[width=0.45\textwidth]{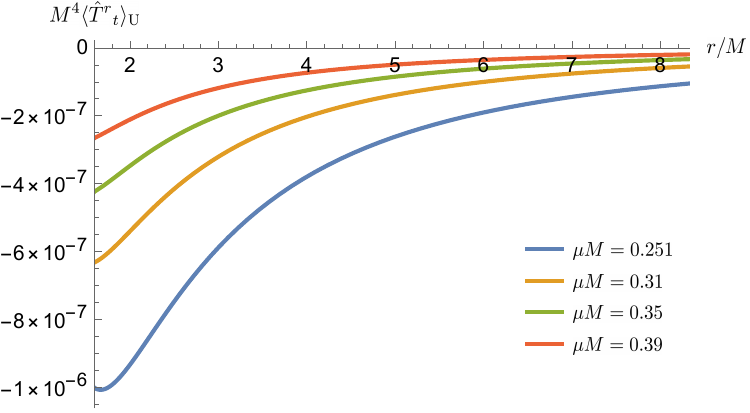}
    \caption{Renormalized expectation values of the charged scalar current (top and middle) and energy flux (bottom) for a charged scalar field on the RN space-time in the Unruh state, with $M=L=1$, $Q/M=0.8$, $\xi=0$, $q M=1/2$ and the scalar mass taking values $\mu M = 0.251$ (blue), $0.31$ (orange), $0.35$ (green), $0.39$ (red).}
    \label{fig:FluxesDifferentMu}
\end{figure}

\begin{figure*}
     \centering
         \includegraphics[width=0.45\textwidth]{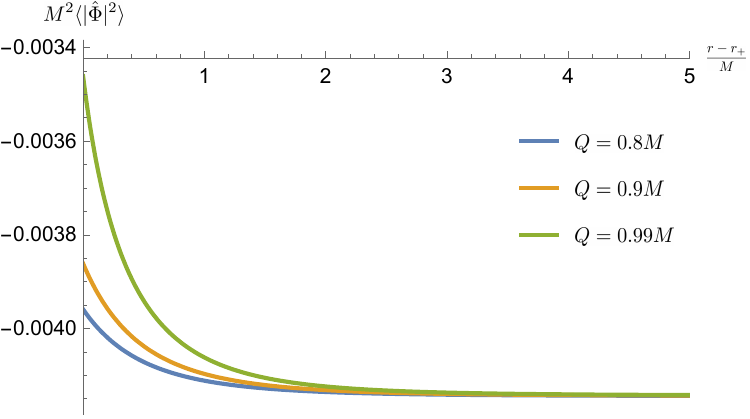}
         \includegraphics[width=0.45\textwidth]{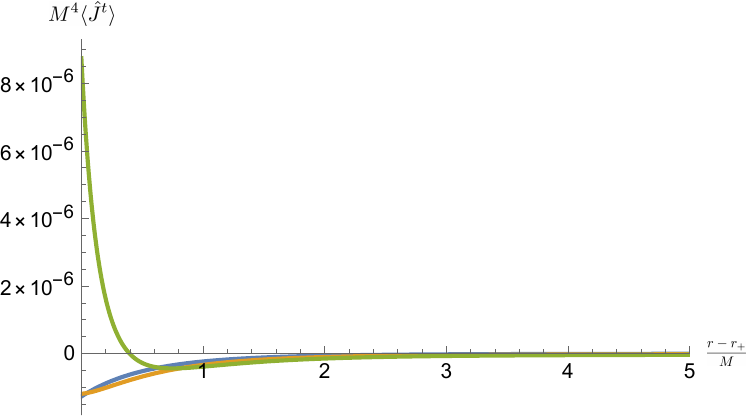}
         \includegraphics[width=0.45\textwidth]{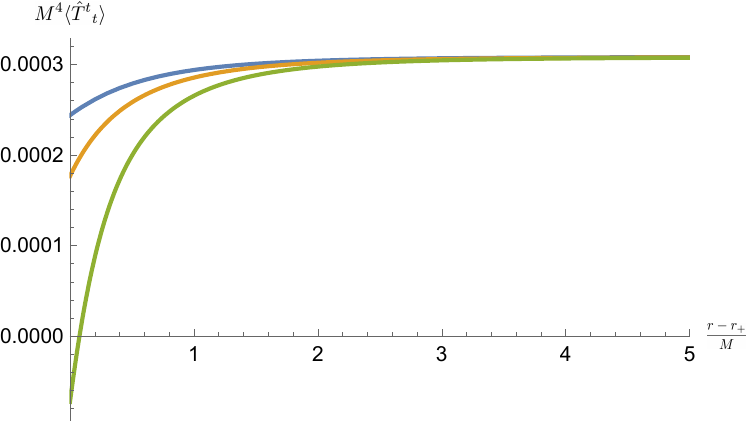}
         \includegraphics[width=0.45\textwidth]{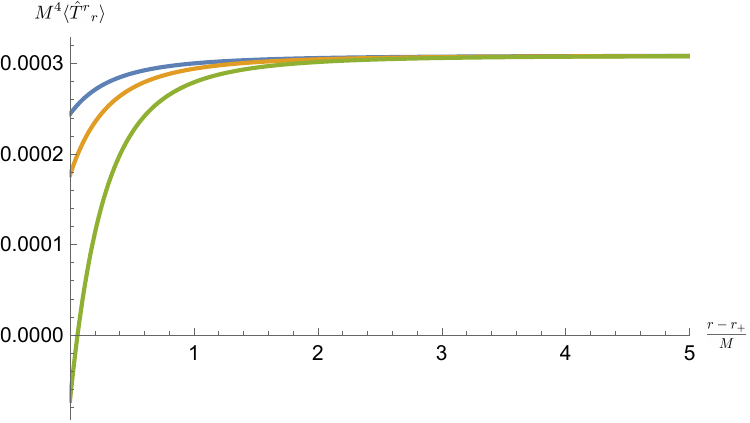}
         \includegraphics[width=0.45\textwidth]{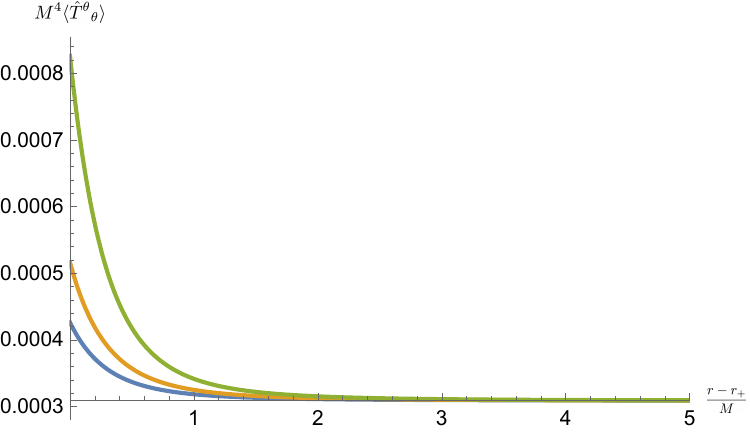}
     \caption{Renormalized expectation values of all observables for a charged scalar field on the RN space-time in the Hartle-Hawking state, with $M=L=1$, $qM=1/2$, $\xi=0$, $\mu M=1/2$ and the black hole charge taking values  $Q/M=0.8$ (blue), $0.9$ (orange), $0.99$ (green).}
     \label{fig:ObsDifferentQ}
\end{figure*}

We close our study of the effect on the renormalized expectation values of varying the scalar field parameters in Fig.~\ref{fig:FluxesDifferentMu}, where we consider only the Unruh state and the charged scalar current components and energy flux. 
The magnitudes of all these quantities increase as the scalar field mass decreases and Hawking radiation becomes more significant. 
Furthermore, all these components tend to zero far from the black hole.
From the top plot in Fig.~\ref{fig:FluxesDifferentMu}, we see that the radial component of the charge current $\langle {\hat {J}}^{r}\rangle _{{\mathrm {U}}}$ is negative everywhere outside the horizon, as expected since the black hole is discharging. 
The time component $\langle {\hat {J}}_{t}\rangle _{{\mathrm {U}}}$ is negative very close to the horizon (corresponding to a positive charge density) but positive in most of the space-time exterior to the horizon.  
For a massless scalar field \cite{Montagnon:2025vtk}, while $\langle {\hat {J}}_{t}\rangle _{{\mathrm {U}}}$ is also negative close to the horizon and becomes positive for sufficiently large $r$, the region where $\langle {\hat {J}}_{t}\rangle _{{\mathrm {U}}}$ is negative is larger than for the massive charged scalar field.
In addition, the maximum of $\langle {\hat {J}}_{t}\rangle _{{\mathrm {U}}}$ is proportionately smaller in the massless compared to the massive case (taking into account that the overall magnitude of the charge current is a couple of orders of magnitude larger for a massless charged scalar field). 
At the horizon, we have $\langle {\hat {J}}_{t}\rangle _{{\mathrm {U}}}= \langle {\hat {J}}^{r}\rangle _{{\mathrm {U}}}$ and thus the charge current expectation value is regular at the future horizon (as $U\langle {\hat {J}}_{U}\rangle $ vanishes there) but not at the past horizon (since $V\langle {\hat {J}}_{V}\rangle $, while finite, is nonzero there). 

\subsection{Varying the black hole charge}
\label{sec:BHcharge}

Thus far we have studied how changing the scalar field parameters $\xi $, $qM$ and $\mu M$ affects the renormalized expectation values.
Our final exploration of the parameter space fixes these scalar field parameters ($\xi = 0$, $qM=1/2$, $\mu M=1/2$) and varies the black hole charge $Q$. 

We begin, as previously, by considering the expectation values in the Hartle-Hawking state in Fig.~\ref{fig:ObsDifferentQ}.  
Here we see that changing the black hole charge does not affect the qualitative shape of the profiles of the charged scalar condensate or RSET components, all of which are either monotonically increasing or decreasing as the radial coordinate $r$ increases. 
This is similar to the situation for a neutral scalar field on an RN black hole \cite{Arrechea:2023fas}, where changing the black hole charge also does not significantly change the qualitative nature of the profiles of the RSET components. 

At the horizon, the  values of the charged scalar condensate and $\langle {\hat {T}}^{\theta }{}_{\theta }\rangle _{{\mathrm {H}}}$ increase as $Q/M$ increase; while the values of the RSET components $\langle {\hat {T}}^{t}{}_{t}\rangle _{{\mathrm {H}}}$ and $\langle {\hat {T}}^{r}{}_{r}\rangle _{{\mathrm {H}}}$ on the horizon decrease as $Q/M$ increases.  
When $Q/M$ is sufficiently close to the extremal limit $Q/M\rightarrow 1$, both $\langle {\hat {T}}^{t}{}_{t}\rangle _{{\mathrm {H}}}$ and $\langle {\hat {T}}^{r}{}_{r}\rangle _{{\mathrm {H}}}$ are negative in a region close to the horizon. 

The behaviour of the charge density $\langle {\hat {J}}^{t} \rangle _{{\mathrm {H}}}$ for $Q/M$ close to unity is particularly striking, being positive close to the horizon but negative further away from the black hole.
In contrast, for smaller values of the black hole charge, the charge density is negative everywhere outside the event horizon (as seen in Figs.~\ref{fig:ObsDifferentsq} and \ref{fig:ObsDifferentMu}). 
This change in behaviour when the black hole is close to extremality is not seen for any of the states for a massless scalar field \cite{Montagnon:2025vtk}.

\begin{figure}
    \centering
    \includegraphics[width=0.45\textwidth]{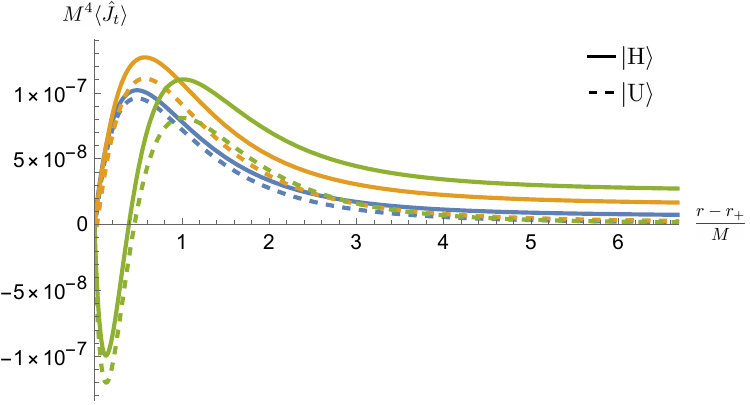}
    \includegraphics[width=0.45\textwidth]{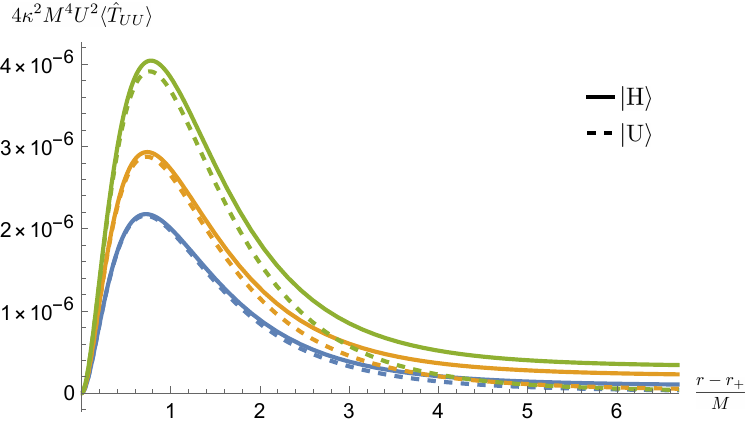}
    \includegraphics[width=0.45\textwidth]{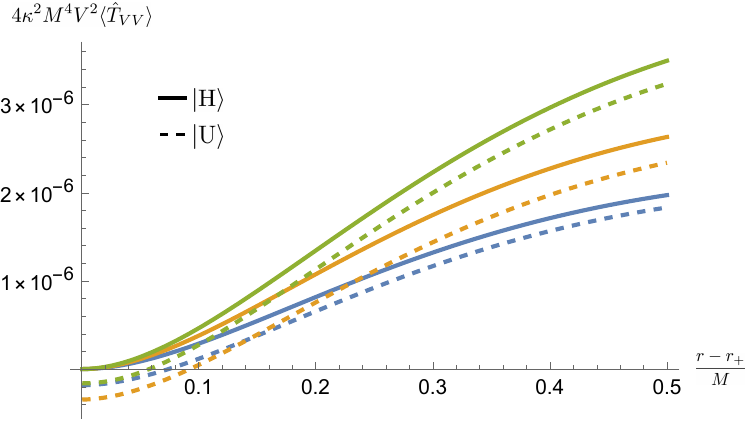}
    \caption{Renormalized expectation values for a charged scalar field on the RN space-time in the Hartle-Hawking and Unruh states, with $M=L=1$, $qM=1/2$, $\xi=0$, $\mu M=1/2$ and the black hole charge taking values  $Q/M=0.8$ (blue), $0.9$ (orange) and $0.99$ (green). Curves for the components of the renormalized expectation values not shown are almost indistinguishable between the two states.}
    \label{fig:ObsDifferentQHU}
\end{figure}

As was the case previously when we varied the scalar field mass and charge, we find that many features of the expectation values in the Hartle-Hawking and Unruh  states are very similar for the parameters used in Fig.~\ref{fig:ObsDifferentQ}.
In Fig.~\ref{fig:ObsDifferentQHU}, we therefore show only those expectation values for which there are visible differences between these two states. 
The top plot in Fig.~\ref{fig:ObsDifferentQHU} shows the component $\langle {\hat {J}}_{t}\rangle $ of the charge current (cf.~$\langle {\hat {J}}^{t}\rangle $ in Fig.~\ref{fig:ObsDifferentQ}), which vanishes on the event horizon in the Hartle-Hawking state but is nonzero in the Unruh state (this is not evident from Fig.~\ref{fig:ObsDifferentQHU}, but can be seen from an inspection of the raw data). 
Close the horizon, this has a different sign for the largest value of $Q/M$ we consider, compared to smaller values of $Q/M$. 
We also see that, far from the black hole, this quantity tends to a smaller value for the Unruh state compared with the Hartle-Hawking state, for fixed $Q/M$.

The remaining plots in Fig.~\ref{fig:ObsDifferentQHU} show the RSET components in Kruskal coordinates (\ref{eq:expKruskal}). 
As before, $U^{2}\langle {\hat {T}}_{UU} \rangle $ vanishes on the horizon in both the Unruh and Hartle-Hawking states, while $V^{2}\langle {\hat {T}}_{VV}\rangle $ is zero on the horizon in the Hartle-Hawking state, but nonzero on  the horizon for the Unruh state. 
While the magnitudes of both quantities generally increase for fixed radial coordinate $r$ as the black hole charge increases, there is no significant qualitative difference as observed for the charge density.

\begin{figure}
    \centering
    \includegraphics[width=0.45\textwidth]{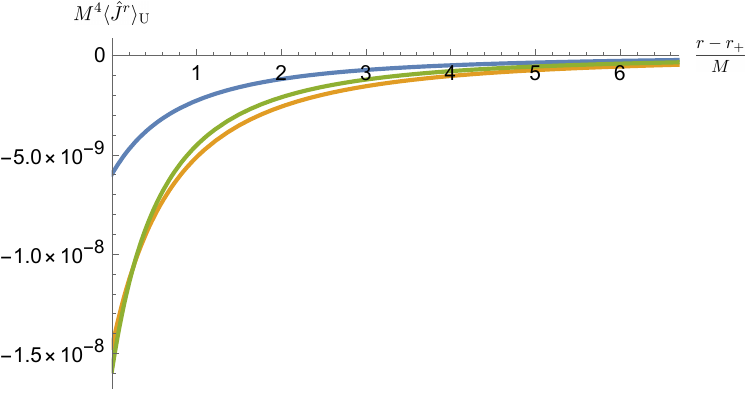}
    \includegraphics[width=0.45\textwidth]{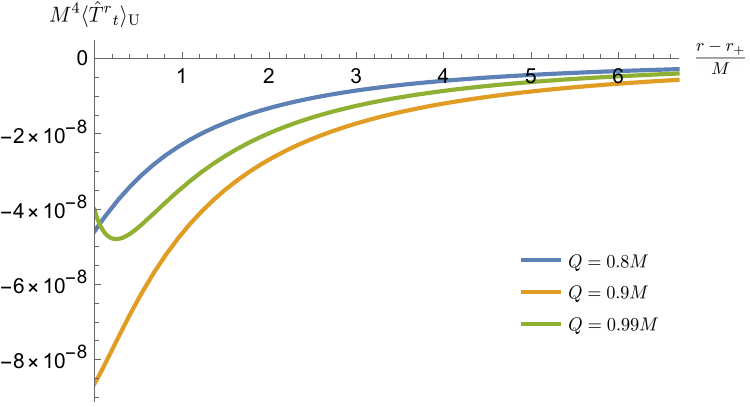}
    \caption{Renormalized expectation values of the charge flux (top) and energy flux (bottom) for a charged scalar field on the RN space-time in the Unruh state, with $M=L=1$, $qM=1/2$, $\xi=0$, $\mu M=1/2$ and the black hole charge taking values  $Q/M=0.8$, $0.9$ and $0.99$.}
    \label{fig:FluxesDifferentQ}
\end{figure}

We close our discussion of the effect of changing the black hole charge by considering, in Fig.~\ref{fig:FluxesDifferentQ}, the fluxes of charge $\langle {\hat {J}}^{r}\rangle _{{\mathrm {U}}}$ and energy $\langle {\hat {T}}^{r}{}_{t}\rangle _{{\mathrm {U}}}$ in the Unruh state. 
As expected, the charge flux is negative for all values of $Q/M$ considered, so that the black hole is losing charge. 
For fixed $r$, the magnitude of $\langle {\hat {J}}^{r}\rangle _{{\mathrm {U}}}$ increases with increasing black hole charge, except when the black hole is very close to extremality.
We find similar behaviour for the energy flux.  
For all $Q/M$, we see that $\langle {\hat {T}}^{r}{}_{t}\rangle _{{\mathrm {U}}}$ is negative, corresponding to the black hole losing energy. 
The magnitude of this quantity increases as $Q/M$ increases from $0.8$ to $0.9$, but decreases as $Q/M$ increases further from $0.9$ to $0.99$.

\section{Conclusions}
\label{sec:conc}

In this work we have presented numerical results for the renormalized expectation values of the scalar condensate, charge current and stress energy tensor (RSET), for a massive, charged quantum scalar field in the Boulware, Unruh and Hartle-Hawking states, on the exterior region of the RN space-time. 
The results in the Hartle-Hawking state were obtained by utilising the extended coordinate scheme which was adapted in \cite{Breen:2024ggu} for massive, charged quantum scalar fields on static, spherically symmetric space-times. 
This scheme is implemented on the Euclidean section of any such space-time for the field in the Hartle-Hawking state. 
In the context of this paper (where we work in the parameter regime corresponding to the absence of superradiance) we expect that one could alternatively work with the Boulware state on the Euclidean section, extending the work of \cite{Arrechea:2024cnv} to the charged situation analogously to \cite{Breen:2024ggu}. 
However, performing the renormalization calculation in the Hartle-Hawking state is typically less computationally taxing than for the Boulware state, due to the discrete frequency spectrum of the Euclidean Green function.

Having obtained renormalized results in the Hartle-Hawking state, we appealed to the finitude of the difference of expectation values in distinct Hadamard states to derive results for the expectation values of the field in the Unruh and Boulware states. 
Taking the difference of the expectation values of interest in either the Boulware or Unruh states with those in the Hartle-Hawking state results in a convergent integral over Lorentzian modes. 
This was evaluated numerically and combined with the Hartle-Hawking result to obtain the desired expectation values in the Unruh or Boulware states. 
This approach of using a combination of the extended coordinates method and state subtraction to obtain numerical results for renormalization expectation values in a variety of states parallels that of Ref.~\cite{Arrechea:2023fas} for a neutral scalar field. 

The results for the renormalized expectation values allowed us to comprehensively study and compare the physical properties of the Boulware, Unruh and Hartle-Hawking states. 
In particular, the numerical evidence suggests that the states satisfy the same event horizon regularity properties as their neutral counterparts, that is, the Hartle-Hawking state is regular on both the future and past horizons; the Unruh state is regular on the future but not the past horizon; and the Boulware state diverges on both the future and past horizons.  
We were also able to compare our results with those for a massless, charged quantum scalar field in the Unruh, (past) Boulware and CCH states found in \cite{Montagnon:2025vtk}. 
A key difference in the latter case is the presence of charge superradiance \cite{Balakumar:2020gli}. 
The form of the extended coordinates scheme developed in \cite{Breen:2024ggu} for the black hole exterior is only suitable in the regime where there is no charge superradiance, due to its dependence on static states and working on the Euclidean sector. 
Extension of the approach of this paper to encompasses the regime where there are classically  superradiant modes therefore requires further development of the extended coordinates method such that it is applicable for the field in nonequilibrium states (such as Unruh), or it can be implemented directly on the Lorentzian sector. 
One could alternatively use a different renormalization scheme such as the pragmatic mode-sum scheme used in \cite{Montagnon:2025vtk}, however this is more computationally expensive and has yet to be applied to the RSET for a charged scalar field. 

The combination of the extended coordinate and state-subtraction schemes is sufficiently numerically efficient to allow for an exploration of large regions of the parameter space, comprising of the scalar field mass and charge, curvature coupling and black hole charge. 
In this paper, we investigated varying these parameters for the field in the Hartle-Hawking and Unruh states. 
In some instances, we found little difference between the results in these two states, however we believe that this is a consequence of considering a comparatively large scalar field mass. 
We also obtained results for parameter values close to the superradiant bound (\ref{eq:bound}) and the black hole extremal limit $Q=M$. 

The results obtained in this paper may be applied to study the backreaction of the quantum field on the space-time metric and the background electromagnetic field, which are governed, respectively, by the semiclassical Einstein (\ref{eq:semieqs}) and Maxwell (\ref{eq:SCMaxwell}) equations. 
Considering the fluxes of charge and mass, solving these equations perturbatively would allow us to investigate how the black hole’s mass and charge evolve through its evaporation process.
This evolution has been studied using an approximation for the fluxes \cite{Hiscock:1990ex,Ong:2019vnv}.
Employing an analytic approximation for the massless charged scalar field modes valid in the near-extremal limit, recent work \cite{Alberti:2025mpg} has shown that the backreaction causes the black hole to evolve away from extremality when the field is in the Unruh state, in agreement with the results of \cite{Hiscock:1990ex,Ong:2019vnv}.  
Going beyond the near-extremal limit, the work of \cite{Hiscock:1990ex,Ong:2019vnv} reveals an ``attractor'' in the $(M,Q/M)$ plane.
It would be of interest to see whether this feature persists when the exact quantum fluxes for a charged scalar field are incorporated into the evolution equations. 

While studies of the backreaction of a charged quantum scalar field to date have largely focussed on the effect of the fluxes on the black hole mass and charge \cite{Klein:2021ctt,Alberti:2025mpg}, the semiclassical Einstein (\ref{eq:semieqs}) and Maxwell (\ref{eq:SCMaxwell}) equations govern the effect of the quantum field on all components of the metric and electromagnetic field, as functions of the space-time coordinates.
Studying the backreaction beyond the effect of the fluxes is a challenging problem which has received little attention in the literature. 
One notable recent result is in Ref.~\cite{Montagnon:2025vtk} where, through studying the semiclassical Maxwell equations (\ref{eq:SCMaxwell}) only, the authors found the possibility of a generic logarithmic divergence in the quantum-corrected electromagnetic field for nonequilibrium  states (such as the Unruh state studied in this paper). 
However, for the metric ansatz used in \cite{Montagnon:2025vtk}, the semiclassical Maxwell equations involve a metric perturbation which can only be determined by solving the semiclassical Einstein equations, for which the full RSET is required.
Therefore, it is possible that a full analysis of the backreaction, encompassing both sets of equations, may rectify this issue. 
In order to explore this question, an efficient renormalization scheme (such as presented here \cite{Breen:2024ggu}) is required to compute accurate results across a wide range of parameters.
Furthermore, our methodology makes possible an investigation of the backreaction effects of the quantum field in different states. 
We will investigate the backreaction problem in our upcoming work \cite{Breen:2026}.

Finally, we anticipate that the approach of this paper could be extended to the black hole interior (at least for the Hartle-Hawking and Unruh states which are defined there), which has been the focus of recent work on the charged scalar field fluxes \cite{Klein:2021ctt,Alberti:2025mpg}. 
The extended coordinate method is applicable on the region of the Lorentzian space-time bounded by the event and Cauchy horizon due to the positive definite nature of the metric on fixed $r$ hypersurfaces.  
As one works on the Lorentzian section, one has a continuous frequency spectrum in the decomposition of the Green function that is used for the calculation of the expectation values. 
One would therefore have to use the Boulware extended coordinate methodology of \cite{Arrechea:2024cnv}, however, as discussed above, we envision no fundamental issues to extending this to the charged case. 
Again, we postpone such investigations to future work.

\begin{acknowledgments}
We acknowledge IT Services at The University of Sheffield for the provision of services for High Performance Computing. The work of E.W.~is supported by STFC grant number ST/X000621/1.
\end{acknowledgments}

\section*{Data availability}
The data that support the findings of this article are
openly available \cite{figshare}.

\bibliography{charge}

\end{document}